%% file: main.tex
\documentclass[conference]{IEEEtran}
\IEEEoverridecommandlockouts

\usepackage[nocompress]{cite}
\usepackage{amsmath,amssymb,amsfonts}
\usepackage[ruled,vlined,linesnumbered]{algorithm2e}
\usepackage{graphicx}
\usepackage{textcomp}
\usepackage{xcolor}
\def\BibTeX{{\rm B\kern-.05em{\sc i\kern-.025em b}\kern-.08em
    T\kern-.1667em\lower.7ex\hbox{E}\kern-.125emX}} 

\usepackage[utf8]{inputenc} %
\usepackage[T1]{fontenc}    %
\usepackage{hyperref}       %
\usepackage{url}            %
\usepackage{booktabs}       %
\usepackage{xcolor}         %
\usepackage{amsthm}        %

\theoremstyle{definition}

\theoremstyle{plain}

\theoremstyle{remark}

\usepackage{multirow}
\usepackage{multicol}
\usepackage{graphicx}
\usepackage{subcaption}
\usepackage{makecell}
\usepackage{tcolorbox}

\usepackage{pifont} %
\newcommand{\cmark}{\ding{51}}
\newcommand{\xmark}{\ding{55}}

\tcbuselibrary{skins,breakable}
\newtcolorbox{reviewbox}{
  enhanced,
  boxrule=0.6pt,
  colback=gray!5,
  colframe=gray!60,
  arc=2pt,
  left=6pt,right=6pt,top=4pt,bottom=4pt,
  boxsep=2pt
}
\newtcolorbox{findingbox}{
  enhanced,
  boxrule=0.6pt,
  colback=gray!5,
  colframe=gray!60,
  arc=2pt,
  left=6pt,right=6pt,top=4pt,bottom=4pt,
  boxsep=2pt
}

\newif\ifshowrevisions
\showrevisionsfalse     %

\ifshowrevisions
  \newcommand{\rev}[1]{\textcolor{blue}{#1}}
  \newenvironment{revblock}{%
    \begingroup
    \color{blue}
    \captionsetup{font={color=blue}} %
  }{%
    \endgroup
  }
\else
  \newcommand{\rev}[1]{#1}
  \newenvironment{revblock}{%
    \ignorespaces %
  }{%
    \ignorespacesafterend %
  }
\fi

\usepackage{todonotes}

\title{WikiDBGraph: A Data Management Benchmark Suite for Collaborative Learning over Database Silos
\thanks{*Equal Contribution}}

\author{\IEEEauthorblockN{Zhaomin Wu*}
\IEEEauthorblockA{\textit{National University of Singapore}\\
Singapore\\
zhaomin@u.nus.edu}
\and
\IEEEauthorblockN{Ziyang Wang*}
\IEEEauthorblockA{\textit{National University of Singapore}\\
Singapore\\
wangziyang@u.nus.edu}
\and
\IEEEauthorblockN{Bingsheng He}
\IEEEauthorblockA{\textit{National University of Singapore}\\
Singapore\\
dcsheb@nus.edu.sg}}

\begin{document}

\maketitle

\begin{abstract}
  \input{sections/abstract}

\end{abstract}


\section{Introduction}
\label{sec:introduction}

\input{sections/introduction}

\begin{revblock}
\section{Background}
\label{sec:background}

\input{sections/background}
\end{revblock}

\section{Related Work}
\label{sec:related_work}

\input{sections/related-work}

\section{Dataset Construction}
\label{sec:dataset_construction}

\input{sections/dataset_construction}

\section{Dataset Details}
\label{sec:dataset_details}

\input{sections/dataset_details}

\section{Experiments}
\label{sec:experiments}

\input{sections/experiment}

\section{Case Study}
\label{sec:case_study}

\input{sections/case_study}

\section{Conclusion}
\label{sec:conclusion_broader_impact}

\input{sections/conclusion}

\section*{AI-Generated Content Acknowledgement}
GPT-5 is used for light polishing of human-written content. Claude and GPT-5 assist with code implementation. All contents are finally human-verified.

\section*{Acknowledgement}
This research/project is supported by the National Research Foundation, Singapore and Infocomm
Media Development Authority under its Trust Tech Funding Initiative. Any opinions, findings and
conclusions or recommendations expressed in this material are those of the author(s) and do not
reflect the views of National Research Foundation, Singapore and Infocomm Media Development
Authority.

\bibliographystyle{ieeetr}
\bibliography{references}

\clearpage
\input{tech-report}

\end{document}

%% file: sections/abstract.tex
Relational databases are often fragmented across organizations, creating data silos that hinder distributed data management and mining. Collaborative learning (CL)---techniques that enable multiple parties to train models jointly without sharing raw data---offers a principled approach to this challenge. However, existing CL frameworks (e.g., federated and split learning) remain limited in real-world deployments. Current CL benchmarks and algorithms primarily target the learning step under assumptions of isolated, aligned, and joinable databases, and they typically neglect the end-to-end data management pipeline, especially preprocessing steps such as table joins and data alignment. In contrast, our analysis of the real-world corpus \textit{WikiDBs} shows that databases are interconnected, unaligned, and sometimes unjoinable, exposing a significant gap between CL algorithm design and practical deployment. To close this evaluation gap, we build \textit{WikiDBGraph}, a large-scale dataset constructed from 100{,}000 real-world relational databases linked by 17 million weighted edges. Each node (database) and edge (relationship) is annotated with 13 and 12 properties, respectively, capturing a hybrid of instance- and feature-level overlap across databases. Experiments on \textit{WikiDBGraph} demonstrate both the effectiveness and limitations of existing CL methods under realistic conditions, highlighting previously overlooked gaps in managing real-world data silos and pointing to concrete directions for practical deployment of collaborative learning systems.

%% file: sections/introduction.tex
Relational databases are prevalent in real-world applications \cite{haskin1982extending,codd2007relational}. These databases are often heterogeneously distributed across clients, creating data silos~\cite{kim2016collaborative,li2022federated} that hinder large-scale data management. To overcome this limitation, \textit{collaborative learning} (CL)---a data management paradigm in which multiple parties jointly train models on decentralized data without sharing raw data---has emerged as a promising solution \cite{fu2021vf2boost,fu2022blindfl,wu2020privacy,fu2022towards}.

Existing CL frameworks predominantly focus on the training stage, including federated learning~\cite{li2022federated}, transfer learning~\cite{zhuang2020comprehensive}, and split learning~\cite{vepakomma2018split}. Although CL attains strong performance on established benchmarks, including LEAF~\cite{caldas2018leaf}, OARF~\cite{hu2022oarf}, VertiBench~\cite{wu2023vertibench}, and FedML~\cite{he2020fedml}, production-ready CL systems remain scarce~\cite{daly2024federated}, revealing a significant gap between current CL benchmarks and practical, end-to-end data management systems.

\begin{revblock}
The limitations of existing CL benchmarks are illustrated by a real example from WikiDBs~\cite{vogel2024wikidbs}, a recently released large-scale database corpus: a collaboration between a National Monuments database and a Historic Places Registry (Fig.~\ref{fig:collab}). Standard benchmarks typically approximate such a setting by perfectly partitioning a single table, implicitly assuming ideal schema and instance alignment. In practice, independently curated web databases exhibit structured yet imperfect overlap: some databases share identifiable attributes such as \texttt{Latitude} and \texttt{Longitude}, but differ in schema (e.g., \texttt{ArchitecturalStyle} vs. \texttt{EstablishmentDate}) and overlap only partially at the instance level—some monuments are historic places, but not all. As detailed later in our hybrid-overlap case study (Section VII-D), this specific component bridges two horizontally aligned clusters, enabling practical collaborative tasks such as predicting national-treasure categories for historic places using these partially vertical connections. This example motivates web-based data for CL evaluation: decentralized curation (e.g., via Wikidata) naturally produces autonomous, weakly aligned databases that more faithfully proxy real-world enterprise silos, thereby exposing essential data-management challenges—such as schema matching and fuzzy joining—that synthetic, perfectly aligned benchmarks \cite{li2022federated} typically omit.

\end{revblock}

\begin{revblock}
\begin{figure}[t]
    \centering
    \includegraphics[width=0.95\columnwidth]{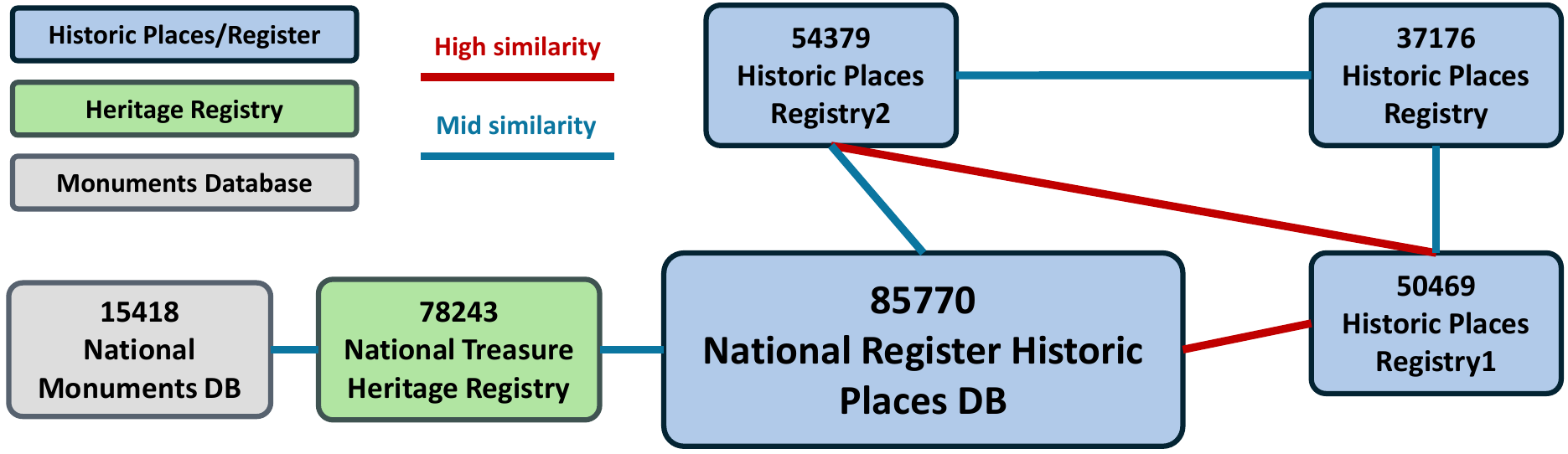}
    \caption{An example of collaborative learning}
    \label{fig:collab}
\end{figure}
\end{revblock}

By investigating WikiDBs~\cite{vogel2024wikidbs}, we identify three common yet impractical assumptions in existing CL benchmarks and algorithm design: (1) \textbf{isolation}: each database is treated as an isolated data source, and connections between clients are ignored; (2) \textbf{alignment}: databases are assumed to be aligned either horizontally (disjoint samples, same features) \cite{yang2019federated} or vertically (disjoint features, same samples) \cite{liu2024vertical}; and (3) \textbf{joinable}: databases can be fully joined into a single database.

Real-world data in WikiDBs~\cite{vogel2024wikidbs} shows the opposite of these assumptions: (1) \textbf{interconnected}: some databases are interconnected according to their schema embeddings while others are not, forming a graph structure; (2) \textbf{unaligned}: databases are rarely aligned horizontally or vertically and are only functionally dependent on each other; even when some databases share the same set of columns, identifying column correspondences is non-trivial due to noisy names and orders; and (3) \textbf{unjoinable}: some databases cannot be fully joined into a single table because of the significant size, which disables most collaborative learning algorithms.

To bridge these gaps and better represent real-world CL tasks, we revisit WikiDBs~\cite{vogel2024wikidbs}. Although it already extracts many relational databases from Wikidata, two challenges impede its direct use for CL: (1) explicit correlations between databases are scarce---only thousands of unconnected database pairs among 100{,}000 databases---which restricts the amount of data available for CL; and (2) database alignment and joinability are not provided, despite being necessary for CL.

We address these challenges by constructing \textbf{WikiDBGraph}, a large-scale, open-source graph of relational databases built from Wikidata~\cite{vrandecic2014wikidata}. To mitigate the limited explicit correlations, we leverage known relations within WikiDBs~\cite{vogel2024wikidbs} and employ contrastive learning to train a model that predicts similarity between databases. The resulting similarity scores induce a graph whose nodes represent databases and whose weighted edges quantify their correlation. To address unaligned and unjoinable databases, we enrich the graph by defining and computing a set of properties for both nodes and edges, derived from database structure, schemas, and data distributions. Using these properties, we propose an automated data-mining pipeline for CL. While this pipeline yields average performance improvements over training on isolated databases, it reveals a critical performance gap compared to ideal centralized training. The contributions of this work include:
\begin{enumerate}
    \item We construct \textbf{WikiDBGraph}, a large-scale graph of 100{,}000 interconnected relational databases, \rev{annotated with 13 node properties and 12 edge properties derived from database structure,} to serve as a real-world benchmark for collaborative learning.
    \item We design an automated CL pipeline and evaluate the overall improvement of mainstream CL algorithms on WikiDBGraph-correlated databases, validating the utility of WikiDBGraph.
    \item We conduct case studies on feature-, instance-, and hybrid-overlapped database pairs, summarizing key challenges to guide future research. 
\end{enumerate}
WikiDBGraph dataset \cite{wikidbgraph_data} and all codes \cite{wikidbgraph_code} for its construction and evaluation are publicly available.

The paper is organized as follows. Section~\ref{sec:background} introduces background and motivation of WikiDBGraph; Section~\ref{sec:related_work} surveys database corpora and CL datasets; Section~\ref{sec:dataset_construction} details the graph construction method; Section~\ref{sec:dataset_details} analyzes graph structure and node/edge properties; Section~\ref{sec:experiments} evaluates the benchmark utility and downstream task performance; Section~\ref{sec:case_study} presents feature-, instance-, and hybrid-overlap CL case studies. Section~\ref{sec:conclusion_broader_impact} concludes the paper. For more details, this technical report also includes the extension to GitTables~\cite{hulsebos2023gittables} (Section~\ref{sec:gittables_ext}), the performance relation to database size (Section~\ref{sec:perf_vs_size}), findings and open challenges (Section~\ref{sec:discussion}), and future directions (Section~\ref{sec:future}).

%% file: sections/background.tex
Collaborative learning (CL) trains models across clients without sharing raw data, spanning federated~\cite{fu2021vf2boost,fu2022blindfl,fu2022towards,wu2020privacy,wu2022practical,qin2025federated}, transfer~\cite{zhuang2020comprehensive}, and split learning~\cite{vepakomma2018split,wu2022fedsim,wu2024fet}. Federated learning iteratively aggregates client updates~\cite{mcmahan2017communication}, using methods like FedProx~\cite{li2020federated} and FedOV~\cite{diao2023towards} to mitigate data heterogeneity. When a single global model is insufficient, personalized federated learning~\cite{chen2022personalized,zhang2024personalized} tailors models to individual clients. For instance, applying SFL~\cite{chen2022personalized} to Fig.~\ref{fig:collab} allows the National Monuments database to personalize its model by aggregating messages from connected registries over the correlation graph.

Beyond federated learning, split learning~\cite{vepakomma2018split} achieves collaboration by partitioning a model across clients (and/or a server) and training it through the exchange of intermediate representations and gradients; inference likewise requires the participating parties. This paradigm is particularly well-suited to feature-partitioned settings, where different clients hold complementary feature subsets. Transfer learning~\cite{zhuang2020comprehensive} provides another collaborative mechanism, often enabling one-shot or low-interaction knowledge transfer: one client trains a model and shares it with another client for adaptation via fine-tuning. While effective in certain communication-constrained scenarios, such one-way transfer typically offers weaker mutual adaptation than multi-round protocols such as federated learning or split learning.

Existing CL systems such as FATE~\cite{liu2021fate} have unified multiple privacy-preserving paradigms---including FL, split learning, and transfer learning---and support diverse models, including linear regression, gradient-boosting decision trees, and support vector machines. Despite this progress, real-world CL deployments remain limited, largely because most algorithms target the model-training stage rather than the end-to-end data management pipeline.

%% file: sections/related-work.tex
This section reviews literature relevant to WikiDBGraph. First, we contrast it with established collaborative learning datasets and benchmarks to underscore its unique capability for modeling complex inter-client relationships, as summarized in Table~\ref{tab:fl_comparison}. Second, we compare WikiDBGraph with existing table and database corpora to highlight its structural novelty, with a summary provided in Table~\ref{tab:dataset_comparison}.

\paragraph{Collaborative Learning Benchmarks}
Existing collaborative learning (CL) benchmarks for federated learning (FL) and split learning (SL) often inadequately capture real-world, database-centric collaboration. Many are synthetic, formed by partitioning a centralized dataset to simulate distribution (e.g., FedNoisy~\cite{liang2023fednoisy}, FPLLib~\cite{zhang2023pfllib}, MarsFL~\cite{huang2024federated}, NIIDBench~\cite{li2022federated}, SLPerf~\cite{hu2025slperf}, VertiBench~\cite{wu2023vertibench}), and thus may not reflect natural heterogeneity. Others are built from real-world sources (e.g., LEAF~\cite{caldas2018leaf}, ORAF~\cite{hu2022oarf}, FedML~\cite{he2020fedml}, NUS-WIDE~\cite{chua2009nus}) but assume perfect alignment into horizontal CL (disjoint samples, same features) or vertical CL (disjoint features, same samples)~\cite{yang2019federated}, whereas practical collaboration often involves partial overlap in both samples and features. Moreover, most benchmarks (i) model each client as a single table and (ii) implicitly treat all inter-client relationships as equally relevant, which is unrealistic for multi-table databases and heterogeneous cross-client correlations. Built from real-world Wikidata, WikiDBGraph is designed to capture these practical complexities by supporting hybrid relationships and varying degrees of correlation across clients.

\begin{table*}[tb]
    \caption{Comparison with existing collaborative learning datasets and benchmarks}
    \label{tab:fl_comparison}
    \begin{minipage}{\textwidth}
        \begin{center}
            \begin{tabular}{l c c c c} 
                \toprule
                \textbf{Dataset} & \textbf{Data Partition}\textsuperscript{1} & \textbf{Alignment across Clients}\textsuperscript{2} & \textbf{Clients' Data Format}\textsuperscript{3} & \textbf{Clients' Relationship}\textsuperscript{4} \\
                \midrule
                FedNoisy \cite{liang2023fednoisy} & \multirow{6}{*}{Synthetic} & Horizontal & Table & \xmark \\
                FPLLib \cite{zhang2023pfllib} & & Horizontal & Table & \xmark \\
                MarsFL \cite{huang2024federated} & & Horizontal & Table & \xmark \\
                NIIDBench \cite{li2022federated} &  & Horizontal & Table & \xmark \\
                SLPerf \cite{hu2025slperf} & & Horizontal & Table & \xmark \\
                VertiBench \cite{wu2023vertibench} & & Vertical & Table & \xmark \\
                VFLAIR \cite{zou2025vflair} & & Vertical & Table & \xmark \\
                \midrule
                LEAF \cite{caldas2018leaf} & \multirow{4}{*}{Real} & Horizontal & Table & \xmark \\
                NUS-WIDE \cite{chua2009nus} & & Vertical & Table & \xmark \\
                ORAF \cite{hu2022oarf} & & Horizontal or Vertical & Table & \xmark \\
                FedML \cite{he2020fedml} & & Horizontal or Vertical & Table & \xmark \\
                \midrule 
                \textbf{WikiDBGraph} & \textbf{Real} & \textbf{Horizontal and Vertical} & \textbf{Database} & \cmark \\
                \bottomrule
            \end{tabular}
        \end{center}
        
        \vspace{1ex}
        
        \textsuperscript{1}How is the data distributed across clients? Synthetic: manually split from centralized dataset; Real: naturally distributed across clients. \\
        \textsuperscript{2}How is the data aligned across clients? Horizontal: disjoint samples, same features; Vertical: disjoint features, same samples. \\
        \textsuperscript{3}What is the format of the clients' data? Table: Client data is represented as individual tables; Database: Client data is represented as relational databases. \\
        \textsuperscript{4}Whether there are explicit relationships between clients.
    \end{minipage}
\end{table*}

\begin{table}[ht]
    \centering
    \caption{Comparison of existing real-world table or database (DB) corpora with WikiDBGraph.}
    \label{tab:dataset_comparison}
    \begin{minipage}{\linewidth} %
        \setlength{\tabcolsep}{4pt}
        \begin{tabular}{l c c c c} 
            \toprule
            \textbf{Dataset} & \textbf{Category} & \textbf{DB Corr.}\textsuperscript{1} & \textbf{\#DBs}\textsuperscript{2} & \textbf{\#Tables}\textsuperscript{2} \\
            \midrule
            SQLShare \cite{jain2016sqlshare} & \multirow{2}{*}{Data-Only} & \xmark & - & 3.9k \\
            GitTables \cite{hulsebos2023gittables} & & \xmark & - & 1M \\
            \midrule
            SchemaDB~\cite{christopher2021schemadb} & \multirow{2}{*}{Schema-Only} & \xmark & 2.5k & - \\
            GitSchemas \cite{dohmen2024gitschemas} & & \xmark & 50k & 300k \\
            \midrule
            BIRD \cite{li2023can} & \multirow{5}{*}{Schema-Data} & \xmark & 95 & 694 \\
            CTU Prague \cite{motl2015ctu} & & \xmark & 83 & 813 \\
            Spider \cite{yu2018spider} & & \xmark & 200 & 1020 \\
            SchemaPile \cite{dohmen2024schemapile} & & \xmark & 75.6k & 347k \\
            WikiDBs \cite{vogel2024wikidbs} & & \xmark & 100k & 1.6M \\
            \midrule 
            \textbf{WikiDBGraph (ours)} & \textbf{Schema-Data} & \cmark & \textbf{100k} & \textbf{1.6M} \\
            \bottomrule
        \end{tabular}
        
        \vspace{1ex}
        
        \textsuperscript{1}The correlation between databases. \\
        \textsuperscript{2}Only the number of databases or tables with available data are counted. \\
    \end{minipage}
\end{table}

\paragraph{Database and Table Corpora}
Early tabular corpora primarily consist of isolated tables (e.g., GitTables~\cite{hulsebos2023gittables}), which do not capture the relational structure inherent to real-world databases. To better reflect practice, database corpora with multiple interrelated tables have emerged, spanning \textit{schema-only} resources (e.g., SchemaDB~\cite{christopher2021schemadb}, GitSchemas~\cite{dohmen2024gitschemas}), \textit{data-only} resources (e.g., SQLShare~\cite{jain2016sqlshare}), and \textit{schema-data} corpora. However, existing \textit{schema-data} corpora are often small-scale (e.g., the CTU Prague Relational Learning Repository~\cite{motl2015ctu}) or tailored to specific tasks (e.g., Text-to-SQL benchmarks such as Spider~\cite{yu2018spider} and BIRD~\cite{li2023can}), limiting their utility for general-purpose representation learning.Despite their scale, recent \textit{schema-data} corpora such as SchemaPile~\cite{dohmen2024schemapile} and WikiDBs~\cite{vogel2024wikidbs} largely lack explicit relationships \emph{between} databases, which hinders studying cross-database collaborative learning. WikiDBGraph addresses this gap by systematically constructing inter-database connections, enabling large-scale CL across distinct databases.

%% file: sections/dataset_construction.tex
\begin{figure*}[ht]
    \centering
    \includegraphics[width=0.8\textwidth]{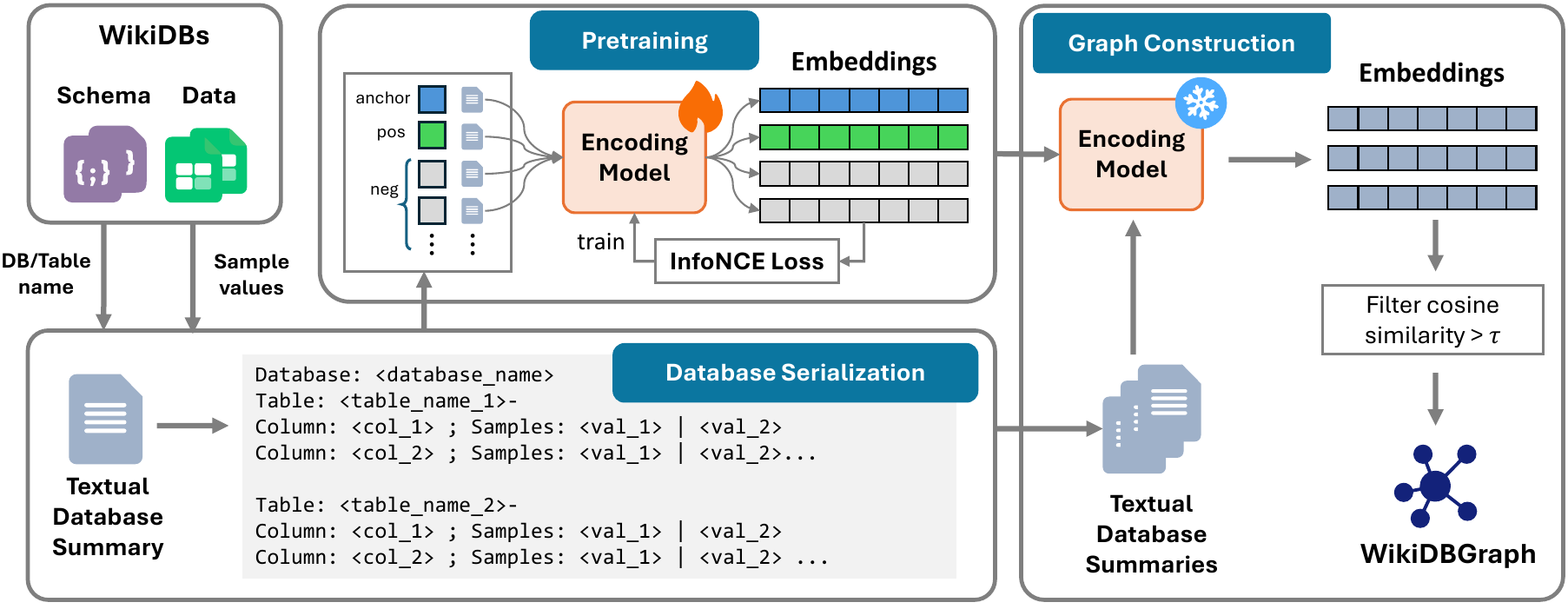}
    \caption{The overview of WikiDBGraph construction process}
    \label{fig:overview}
\end{figure*}

This section details the construction of WikiDBGraph (Fig.~\ref{fig:overview}). We define the database relationship identification task (Section~\ref{subsec:problem-definition}), present our approach (Section~\ref{subsec:approach}), evaluate the embedding model (Section~\ref{subsec:eval-embedding-model}), and describe the final graph construction (Section~\ref{subsec:graph-construction}).

\subsection{Problem Definition}\label{subsec:problem-definition}

This work leverages WikiDBs~\cite{vogel2024wikidbs}, a large-scale collection of relational databases extracted from Wikidata. We aim to construct WikiDBGraph by identifying and establishing correlations between these databases. The WikiDBs construction methodology is explicitly designed to create databases that are thematically coherent, with each one generated from a single starting \textit{topic} from Wikidata. This makes the topic identifier the most direct and high-confidence signal for semantic linkage. We therefore initially define two databases as \textit{correlated} if they share the same value for this foundational attribute, \texttt{wikidata\_topic\_item\_id}, referred to as TID. This approach hypothesizes that two databases seeded from the exact same Wikidata topic are correlated. However, this explicit linkage based on TID identifies only 8,816 correlated pairs among 9,895 distinct databases. This results in a very limited set of connections, leaving over 90\% of the databases without this explicit TID-based link. Such sparsity likely arises because many topics, though intrinsically related, have distinct TIDs within Wikidata.

The primary objective of this work is to train a model capable of uncovering these implicit correlations. Let $\mathcal{D} = \{D_1, D_2, \dots, D_N\}$ and $\mathcal{S} = \{s_1, s_2, \dots, s_N\}$ represent the set of all $N$ databases and $N$ schemas within WikiDBs, respectively, and let 
\begin{equation}
\mathcal{P}_{\text{explicit}} = \left\{(D_i, D_j)|\mathrm{TID}(D_i) = \mathrm{TID}(D_j), i \neq j\right\}
\end{equation}
denote the limited set of explicitly correlated database pairs identified through shared TIDs. We aim to train a function $f: \mathcal{D} \times \mathcal{D} \rightarrow [0, 1]$, where $f(D_i, D_j)$ outputs the predicted probability of a correlation between databases $D_i$ and $D_j$. We model this as a semi-supervised learning problem with limited positive labels in $\mathcal{P}_{\text{explicit}}$.

\subsection{Approach}\label{subsec:approach}

\paragraph{Database Serialization}
To mitigate the verbosity of the original JSON files and reduce input size for subsequent processing, we serialize each $s_i$ and samples in $D_i$ into a concise textual format, denoted as \textit{abstract} $t_i\in\mathcal{T}$. This serialization retains both structural information and an abstract of the data content. Specifically, for each database, we preserve its name, the names of its tables, and for each table, \rev{the names of its columns along with $\phi$ representative sample values (default $\phi=3$) from each column.} An illustrative example of our serialization format is presented below:

{\scriptsize
\begin{verbatim}
Database: <database_name>
Table: <table_name_1>
- Column: <column_name_1> ; Samples: <value_1> | <value_2> 
- Column: <column_name_2> ; Samples: <value_1> | <value_2>
...
Table: <table_name_2>
- Column: <column_name_1> ; Samples: <value_1> | <value_2>
- Column: <column_name_2> ; Samples: <value_1> | <value_2> 
...
\end{verbatim}
}

While sample values are included to offer a qualitative indication of the data, we do not incorporate full data distributions or the entirety of the raw data. This approach is adopted for two primary reasons: 1) schema-level information (database, table, and column names) combined with data samples is often more directly indicative of the database's topic than comprehensive statistical distributions of all values. 2) Processing and representing the complete data for all databases would be computationally expensive and lead to excessively large serialized representations. \rev{While our manual inspection confirms that cryptic metadata (e.g., \texttt{column\_1}) is rare in WikiDBGraph due to upstream curation~\cite{vogel2024wikidbs}, we acknowledge this remains a challenge in broader settings. In such scenarios, our graph formulation offers a natural mitigation: the semantics of opaque schemas can be inferred from the context of well-defined neighboring databases. We leave the empirical evaluation of this resilience for future work.}

\paragraph{Training}
We employ a contrastive learning~\cite{chen2020simple} framework to train an embedding model, $f_\theta(\cdot): \mathcal{T} \rightarrow \mathbb{R}^d$, which maps serialized database schema from the input space $\mathcal{T}$ into a $d$-dimensional vector space. The distinction between positive and negative pairs, crucial for the contrastive loss, is determined by TIDs: a pair of database abstracts $(t_i, t_j)$ is a positive pair if they share the same TID; otherwise, it is a negative pair. The sets of positive and negative pairs are partitioned into training, validation, and test subsets in a 7:1:2 ratio, ensuring no database overlap between partitions.

\begin{figure*}[tb]
    \centering
    \begin{subfigure}[b]{0.29\textwidth}
        \includegraphics[width=\linewidth]{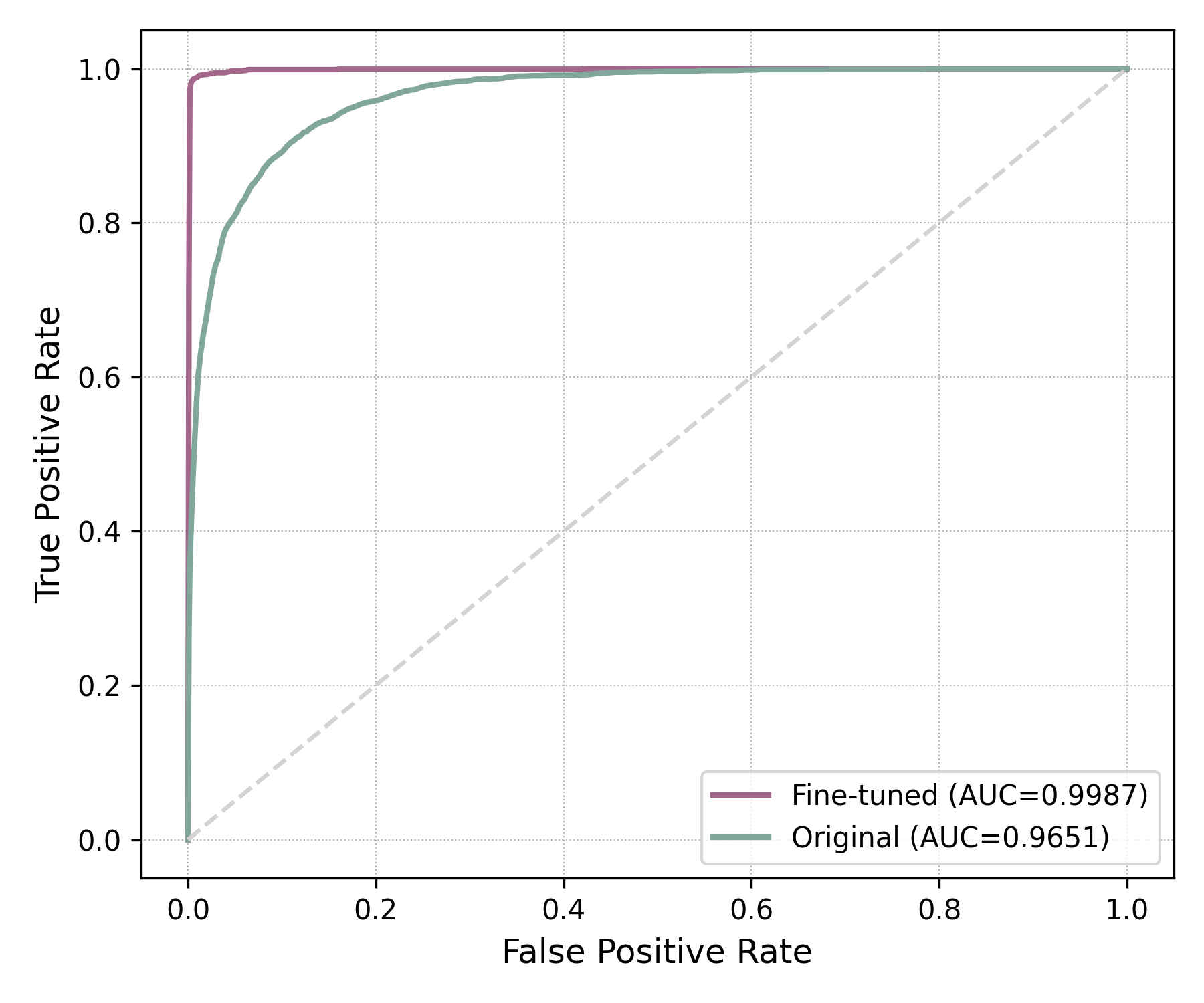}
        \caption{Test set ROC curve}
        \label{fig:roc-curve}
    \end{subfigure}
    \hfill
    \begin{subfigure}[b]{0.29\textwidth}
        \includegraphics[width=\linewidth]{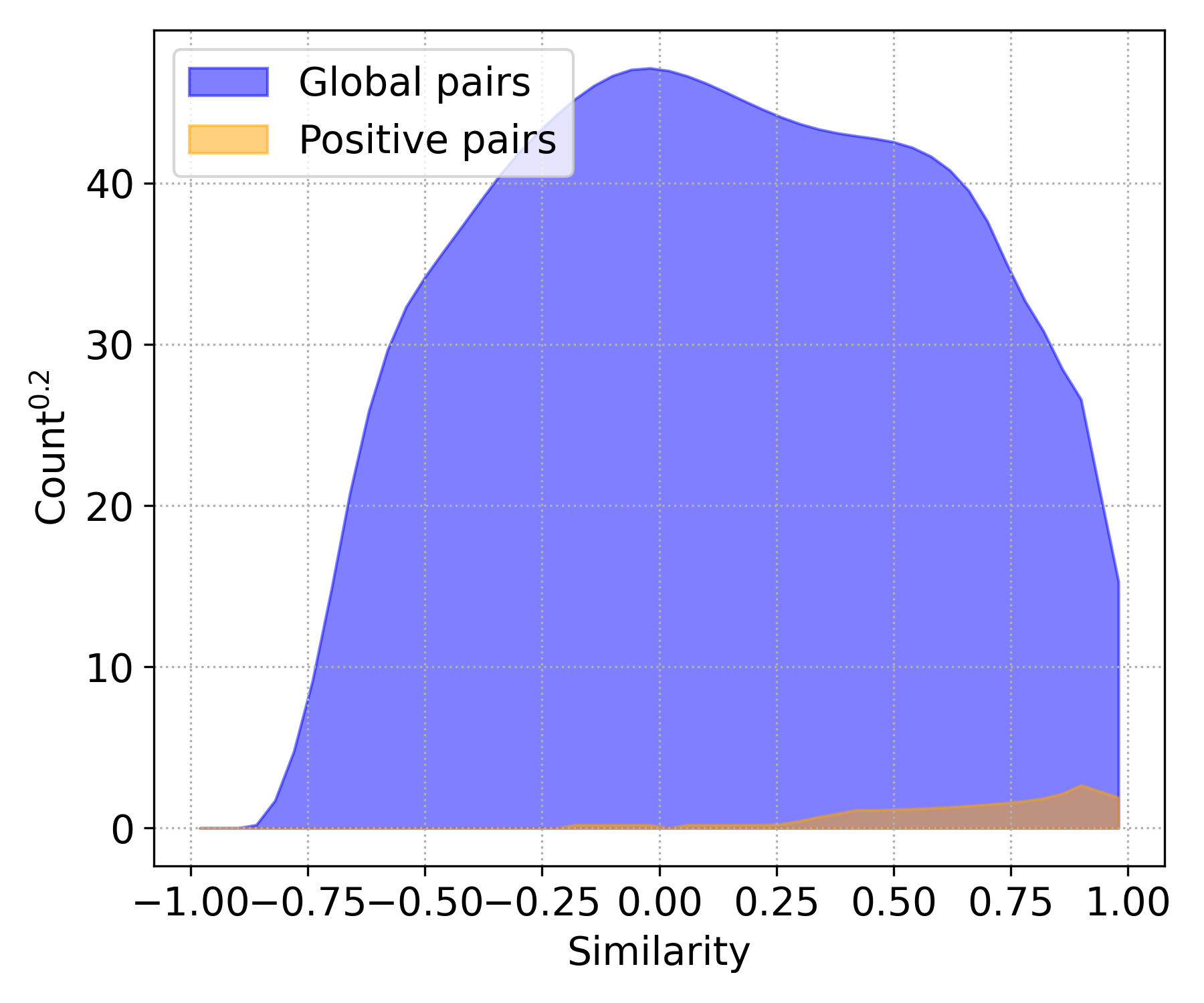}
        \caption{Distribution of similarities}
        \label{fig:dist-pairs}
    \end{subfigure}
    \hfill
    \begin{subfigure}[b]{0.29\textwidth}
        \includegraphics[width=\linewidth]{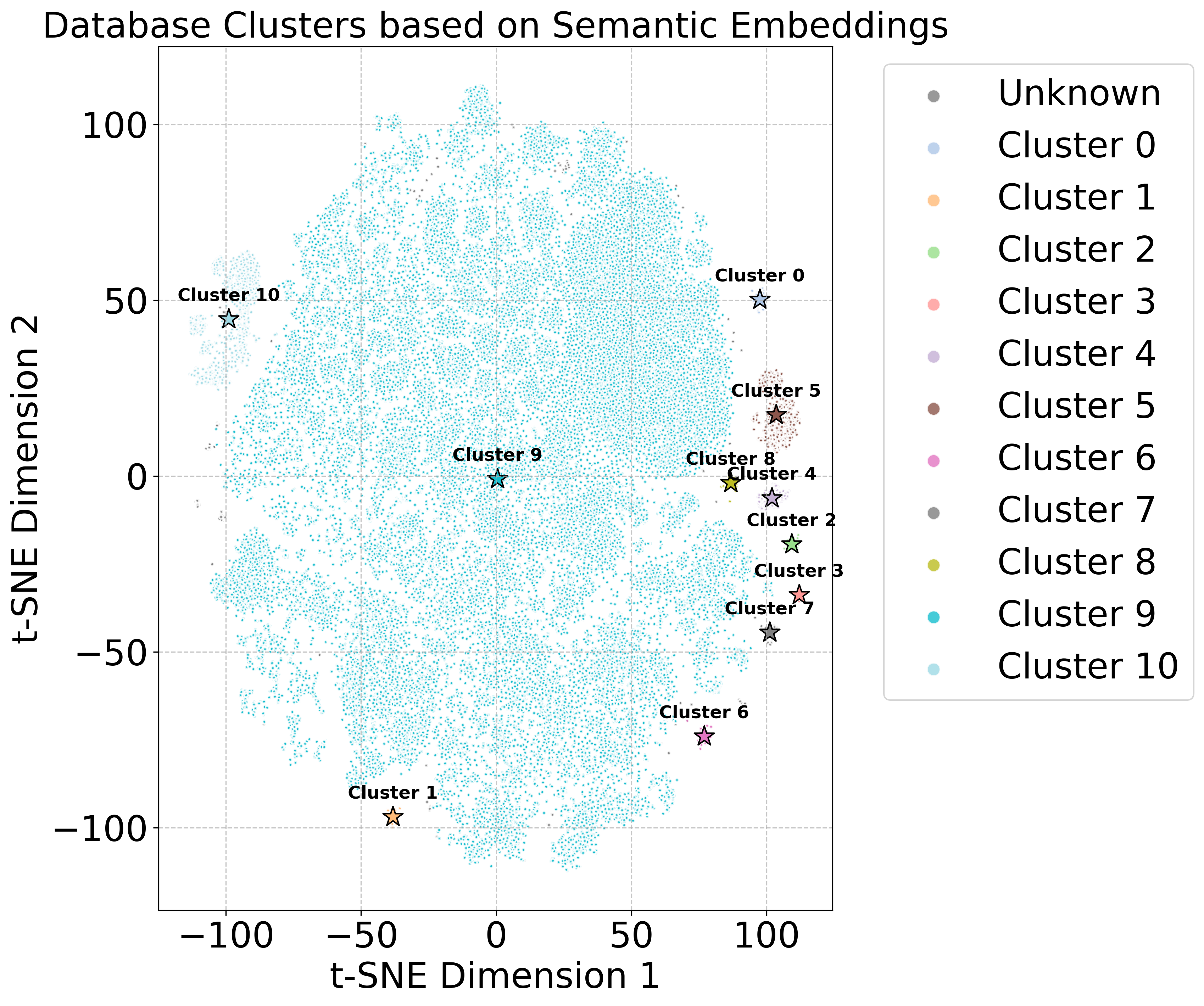}
        \caption{Database HDBSCAN clustering}
        \label{fig:database-clustering}
    \end{subfigure}
    \caption{Performance evaluation of the embedding model}
\end{figure*}

Specifically, the embedding model $f_\theta(\cdot)$ is initialized using parameters from the pretrained encoder-only language model BGE-M3~\cite{chen2024bge}. We select BGE-M3 for its high efficiency, strong performance, and widespread adoption in semantic retrieval tasks~\cite{team2025kimi,faysse2024colpali,han2025videoespresso}, making it well-suited for identifying topical relationships between databases from their textual representations. The model is subsequently fine-tuned during our training process. To construct training instances, each positive pair $(t_a, t_b)$ from the training set is utilized, where $t_a$ serves as the \textit{anchor} and $t_b$ as the \textit{positive abstract}. For each anchor $t_a$, we sample $k$ \textit{negative abstracts} that have distinct TIDs from $t_a$, denoted as $\{t_{n_j}\}_{j=1}^k$, from the training set. Each training instance consists of a triplet $(t_a, t_b, \{t_{n_j}\}_{j=1}^k)$. We set the number of negative samples $k$ to 6, a value yielding sufficiently high AUC in Figure~\ref{fig:roc-curve}.

The model parameters $\theta$ are optimized by minimizing the InfoNCE loss. Initially, the anchor, positive, and negative abstracts are transformed into their respective embeddings:
\begin{equation}
e_a = f_\theta(t_a), \quad e_b = f_\theta(t_b), \quad e_{n_j} = f_\theta(t_{n_j}) \text{ for } j=1, \ldots, k.
\end{equation}
These embeddings, $e_a, e_b, e_{n_j} \in \mathbb{R}^d$, are then employed to compute the InfoNCE loss function~\cite{chen2020simple} as follows:
\begin{equation} \label{eq:infonce}
\mathcal{L}_{\text{InfoNCE}} = -\log \frac{\exp\left(\frac{\mathrm{sim}(e_a, e_b)}{T}\right)}{\exp\left(\frac{\mathrm{sim}(e_a, e_b)}{T}\right) + \sum_{j=1}^k \exp\left(\frac{\mathrm{sim}(e_a, e_{n_j})}{T}\right)},
\end{equation}
where $\mathrm{sim}(\cdot, \cdot)$ represents cosine similarity between two embedding vectors, and $T$ is a temperature hyperparameter controlling the sharpness of the distribution. The optimal embedding model $\theta^*$ is obtained by optimizing the InfoNCE loss function.
\begin{equation}\label{eq:infonce_loss}
\theta^* = \arg\min_{\theta} \mathcal{L}_{\text{InfoNCE}}\left(\theta; \{(t_a, t_b, \{t_{n_j}\}_{j=1}^k)\}_{a,b,j=1}^N\right).
\end{equation}

\begin{figure*}[tb]
    \centering
    \begin{subfigure}[b]{0.29\textwidth}
        \centering
        \includegraphics[width=\linewidth]{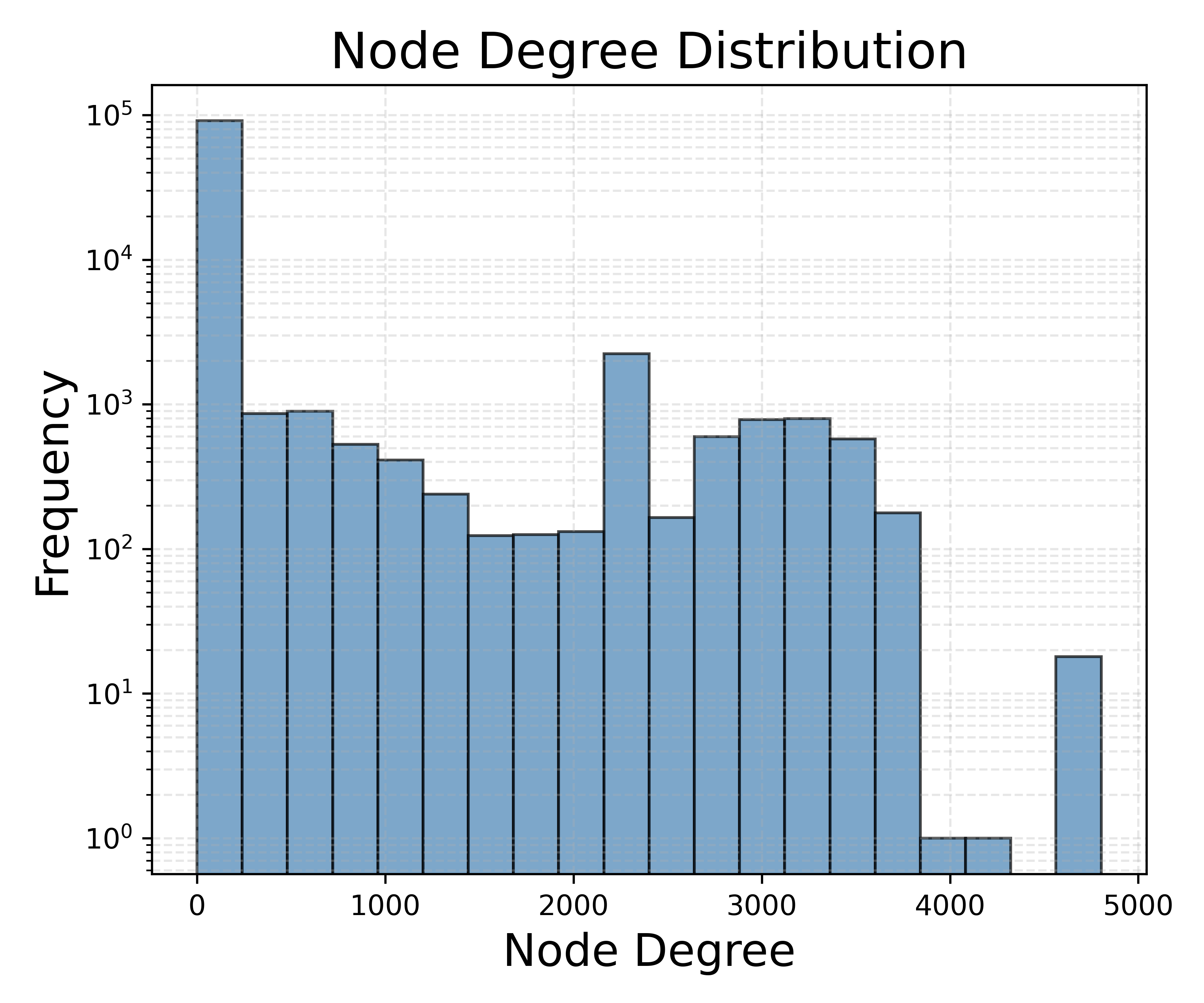}
        \caption{Node Degree}
        \label{fig:degree-distribution}
    \end{subfigure}
    \hfill
    \begin{subfigure}[b]{0.29\textwidth}
        \centering
        \includegraphics[width=\linewidth]{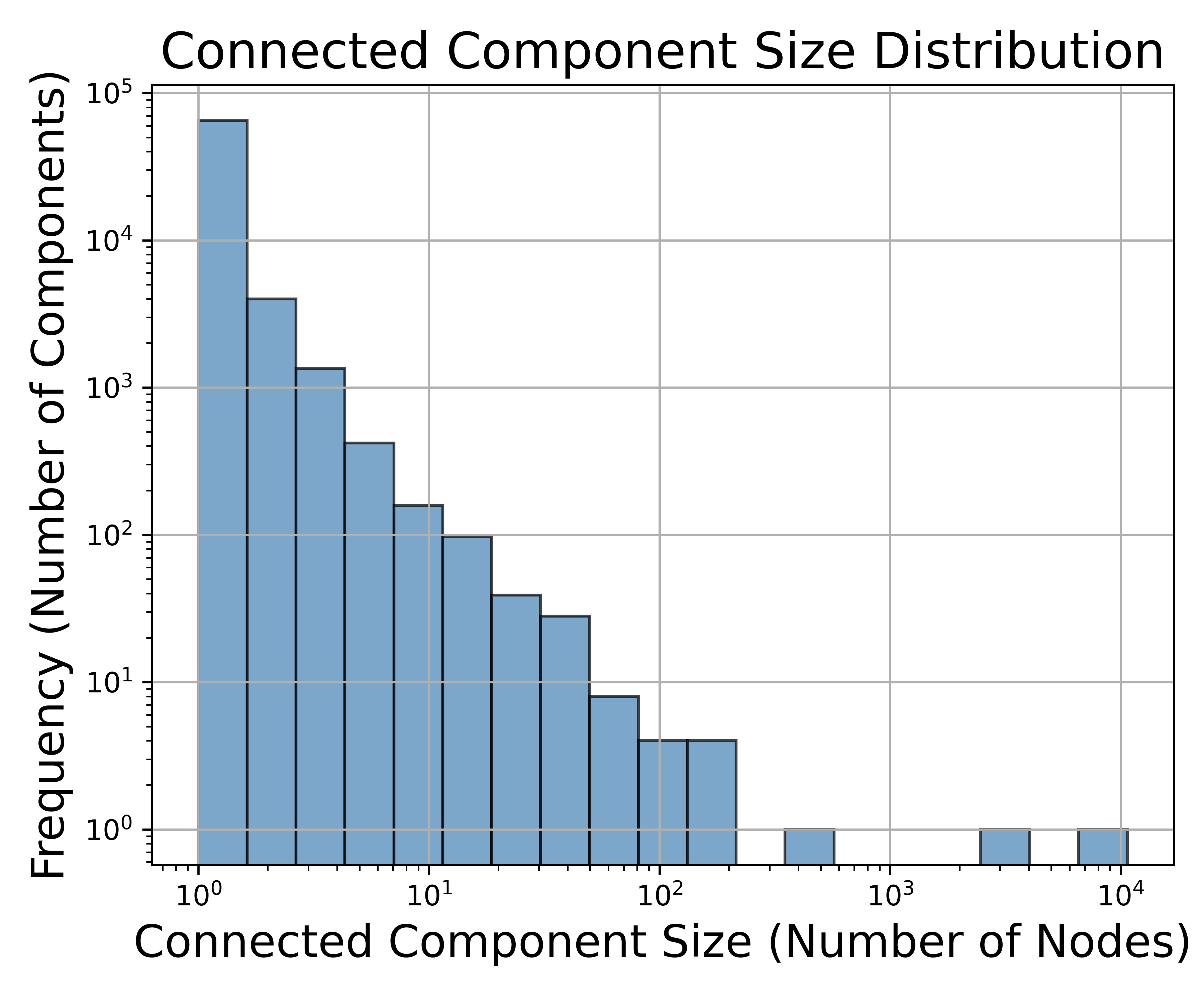}
        \caption{Connected Component Size}
        \label{fig:component-size-distribution}
    \end{subfigure}
    \hfill
    \begin{subfigure}[b]{0.29\textwidth}
        \centering
        \includegraphics[width=\linewidth]{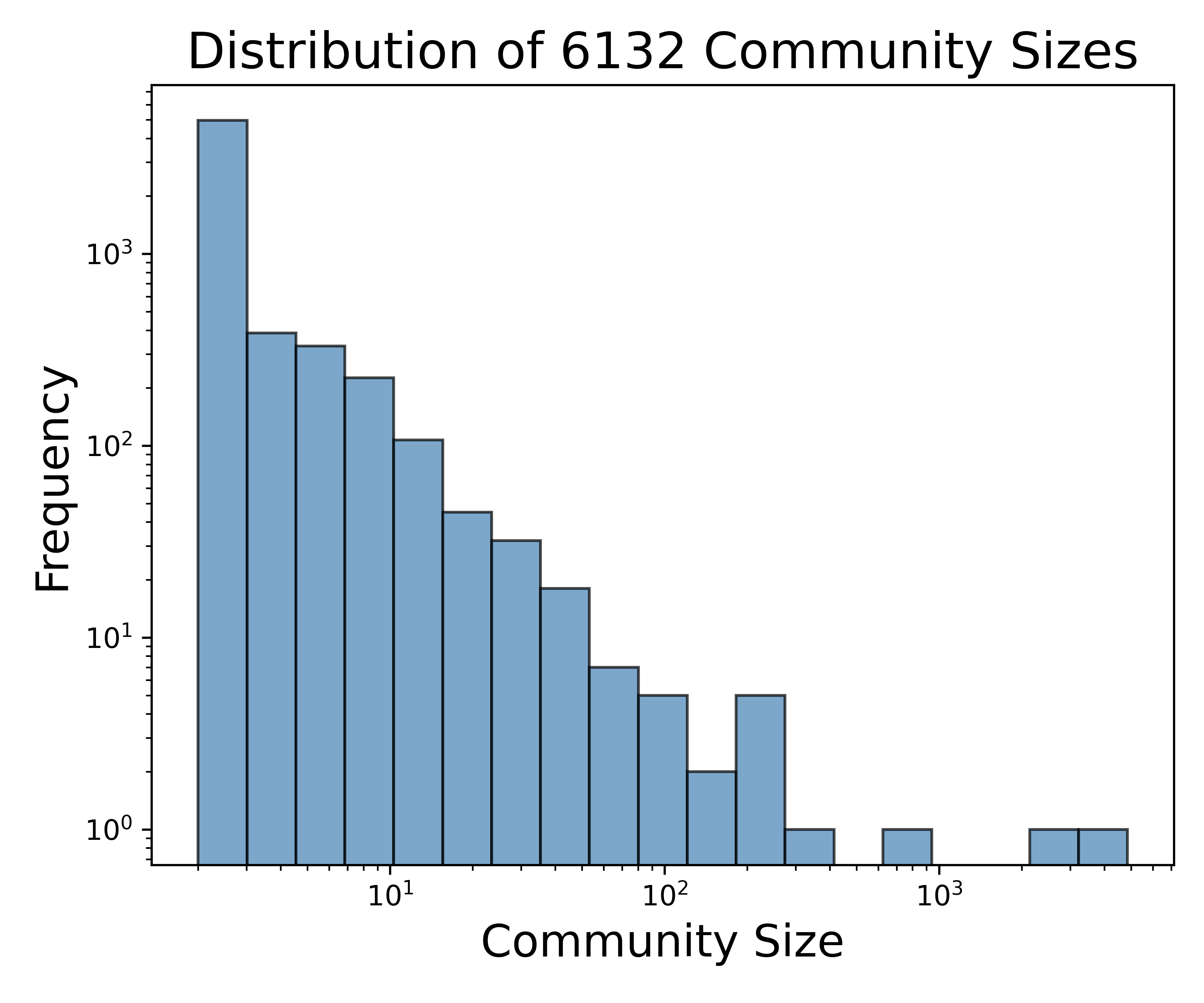}
        \caption{Community Size}
        \label{fig:community-size-distribution}
    \end{subfigure}
    \caption{Distribution of Graph Properties in WikiDBGraph}
    \label{fig:node-properties-distribution}
\end{figure*}

\subsection{Evaluation of Embedding Model}\label{subsec:eval-embedding-model}

\rev{We analyze the similarity distributions in Fig.~\ref{fig:dist-pairs} to validate the reliability of our learned correlations. The distribution of explicit TID-matched pairs is heavily skewed toward $1.0$, confirming that TIDs serve as high-precision labels with negligible false positives. Conversely, the false negatives are very prevalent according to the global distribution revealing a massive tail of high-similarity pairs lacking TID links. Consequently, leveraging these high-precision (low–false-positive) labels, our model can safely expand beyond TID links to uncover a large set of implicit correlations that TID matching misses, substantially enlarging the database network.} Importantly, while these newly discovered correlations are numerous, they remain a small fraction of all possible pairs. Consequently, their discovery does not substantially bias the negative sampling strategy used during training.

The clustering of the database embeddings, projected to two dimensions using t-SNE and subsequently processed with HDBSCAN, is illustrated in Fig.~\ref{fig:database-clustering}. The visualization yields 11 distinct clusters (excluding “unknown”) of varying sizes that correspond to topical categories. The pronounced size imbalance—with Cluster 9 (biomedical) substantially larger than others, e.g., sports (Cluster 10)—is not a clustering artifact but reflects the long-tailed topical distribution of the source corpus, WikiDBs~\cite{vogel2024wikidbs}. WikiDBs contains many biomedical databases (e.g., genes, proteins, diseases), likely due to the prevalence of tabular data in the biomedical domain.

\begin{revblock}

This learned embedding model is not universal across all database corpora. However, the underlying training paradigm readily extends to other corpora, such as GitTables~\cite{hulsebos2023gittables} without pre-defined labels (TIDs). We demonstrate this extension in the technical report~\cite{wikidbgraph_techrpt}.
    
\end{revblock}

\subsection{Graph Construction}\label{subsec:graph-construction}
Upon obtaining the pretrained embedding model, denoted as $f_{\theta^*}$, the embedding vector for each database in the corpus can be derived. \rev{To construct the database graph, we perform nearest-neighbor retrieval in the embedding space and compute cosine similarities between each database embedding and its nearest neighbors. We use the \texttt{faiss} library~\cite{johnson2019billion} to enable efficient large-scale nearest-neighbor search.}

\begin{revblock}

\noindent\textbf{Threshold methodology.}
We treat $\tau$ as a graph sparsification parameter rather than a tuned hyperparameter. We first obtain a data-driven baseline $\tau_0$ from the embedding model’s similarity predictions, using the threshold that maximizes Youden’s index. As shown in Fig.~\ref{fig:threshold_graph}, this choice can produce a very dense graph, which may be undesirable for applications that require efficient construction and analysis. Accordingly, we release (i) all graph-agnostic node/edge properties computed over edges with similarity $\ge \tau_0$, and (ii) the complete graph-construction pipeline, so that users can instantiate graphs under any threshold suited to their budget and use case. In the paper, we additionally report three representative sparsity regimes (low/medium/high) to illustrate how key graph statistics vary with edge pruning (Table~\ref{tab:graph-stats}).

\noindent\textbf{Downstream evaluation and pair selection.}
Our downstream experiments (Section~\ref{sec:experiments}) focus on the highest-similarity pairs because this aligns with how most CL methods are intended to be used: performance is meaningful and comparable only when the paired parties exhibit strong semantic relatedness (e.g., substantial feature/instance overlap). Accordingly, we evaluate on a fixed set of top-ranked pairs ($\tau>0.98$) whose similarities are uniformly very high, making the downstream results effectively insensitive to the particular global sparsification threshold used for graph construction; $\tau$ mainly affects density-related graph analyses and the exploration of more challenging weak-overlap regimes.

\end{revblock}

%% file: sections/dataset_details.tex
This section introduces the structural details of the WikiDBGraph dataset (Section~\ref{subsec:graph-structure}) and subsequently describes the properties defined for the nodes and edges within this graph (Section~\ref{subsec:node-edge-prop}), from which three characteristics emerge.

\subsection{Graph Structure}\label{subsec:graph-structure}

The structural characteristics of the generated database graphs, as detailed in Table~\ref{tab:graph-stats} and further visualized by the distributions presented in Figs.~\ref{fig:degree-distribution}, \ref{fig:component-size-distribution}, and~\ref{fig:community-size-distribution}, reveal the first characteristic of data in CL tasks.

\begin{findingbox}
    \textbf{Characteristic 1:} Client databases form an \textbf{interconnected} graph with non-uniform sparsity and correlation.
\end{findingbox}

The mean degree reported in Table~\ref{tab:graph-stats} is in the hundreds, indicating extensive interconnection among databases. This is further supported by Fig.~\ref{fig:degree-distribution}, which shows a substantial share of high-degree databases. However, the network is heterogeneous: the degree distribution is long-tailed---only a small fraction of databases are high-degree, whereas many reside in small clusters or communities, as evidenced by the connected-component size distribution in Fig.~\ref{fig:component-size-distribution} and the community-size distribution in Fig.~\ref{fig:community-size-distribution}. This imbalance is further visualized in Fig.~\ref{fig:database-clustering}, which reveals many small clusters. Such long-tailed structure with a few high-degree ``hubs'' is common in real-world networks such as social networks~\cite{enders2008long} and webpage networks~\cite{anderson2007long}. Collectively, these results show that the databases in WikiDBGraph are broadly interconnected, yet connectivity varies across databases, underscoring the necessity of a graph structure for CL studies.

\subsection{Node and Edge Properties}\label{subsec:node-edge-prop}

To further characterize data in CL tasks, we compute properties for both nodes (databases) and edges (relationships between databases) in the graph, grouped into structural, semantic, and statistical categories. \textbf{Structural} properties capture architectural characteristics (e.g., table counts and schema similarity). \textbf{Semantic} properties encode conceptual meaning (e.g., topic embeddings and their cosine similarity). \textbf{Statistical} properties quantify the underlying data (e.g., column cardinality and distribution divergence). Tables~\ref{tab:node-props} and \ref{tab:edge-props} summarize these properties; specific details follow.

\paragraph{Nodes} For nodes, which represent individual databases, these properties provide a multifaceted profile that spans structure, semantics, and statistics. \textit{Structural} properties define the database architecture: basic scale metrics include the total number of tables (\texttt{\#Tables}) and columns (\texttt{\#Columns}); \texttt{CategoricalRatio} indicates the proportion of columns containing non-numeric data; and \texttt{ForeignKeyDensity} and \texttt{AverageConnection} quantify internal schema complexity by measuring, respectively, the ratio of foreign keys to columns and the average number of foreign-key links per table. \textit{Semantic} properties capture conceptual identity: each node is characterized by a high-dimensional \texttt{DatabaseEmbed} vector representing its content, its assignment to a broader topic \texttt{ClusterID}, and its \texttt{CommunityID} derived from its position within the graph structure. \textit{Statistical} properties summarize the stored data: \texttt{DataVolume} is the total file size; \texttt{AllJoinSize} estimates the row count resulting from joining all tables; \texttt{AverageCardinality} reflects the average number of unique values per column; \texttt{AverageSparsity} indicates the prevalence of \texttt{NULL} or empty values; and \texttt{AverageEntropy} measures the average information content of the columns.

\paragraph{Edges} Complementing these node-level descriptors, edge properties quantify the nature and strength of relationships between database pairs across structural, semantic, and statistical dimensions. \textit{Structural} properties measure schema similarity via \texttt{JaccardTable}, \texttt{JaccardColumn}, and \texttt{JaccardType} (overlaps of table names, column names, and data types), along with counts of \texttt{CommonTables}, \texttt{CommonColumns}, and \texttt{CommonDataTypes}; \texttt{HellingerDistanceType} captures dissimilarity between data-type distributions, and graph edit distance (\texttt{GED}) quantifies structural differences between schema graphs. \textit{Semantic} properties focus on conceptual likeness: \texttt{EmbedSim} is the cosine similarity between database embeddings, and \texttt{SimilarityConfidence} reflects the certainty of this similarity prediction. \textit{Statistical} properties assess correspondence in shared columns: \texttt{KLDivergence} measures divergence between their data distributions, while \texttt{OverlapRatio} quantifies the proportion of identical values.

These statistics reveal two additional characteristics of data in CL tasks.

\begin{findingbox}
    \textbf{Characteristic 2:} The alignment between databases is a \textbf{hybrid} of horizontal and vertical alignment.
\end{findingbox}

Table~\ref{tab:edge-props} indicates limited \textbf{feature overlap} across databases: the ratios of table and column overlaps are low on average (\texttt{JaccardTable} mean $0.12$; \texttt{JaccardColumn} mean $0.36$), and absolute overlaps remain modest (\texttt{CommonTables} mean $1.85$; \texttt{CommonColumns} mean $74$). Meanwhile, the \textbf{instance overlap} is also limited: the \texttt{OverlapRatio} averages $0.22$, indicating that shared columns contain relatively few identical values across databases. Taken together, these observations imply that most cases cannot be idealized as purely horizontal (same features, disjoint samples) or purely vertical (same samples, disjoint features); rather, the alignment across databases is partial and hybrid.

\begin{findingbox}
    \textbf{Characteristic 3:} Many databases are \textbf{unjoinable} due to the size of the data.
\end{findingbox}

Although the mean \texttt{DataVolume} per database is small (1.5\,MB; Table~\ref{tab:node-props}), the average \texttt{AllJoinSize}, which is the estimated row count from joining all tables within a database, is $9\times 10^{16}$. Such joins are computationally prohibitive in both time and memory, rendering CL methods that assume a single, pre-joined table impractical in this setting. Consequently, practical CL must operate over relational structures directly, avoiding explicit full joins while leveraging graph-aware strategies.

\begin{table*}[tb]
    \begin{center}
        \caption{Summary of Graph Statistical Properties (CC: Connected Component)}
    \label{tab:graph-stats}
    \begin{tabular}{c cccccc cccc cccc}
        \toprule
        \multirow{2}{*}{Confidence\textsuperscript{1}} & \multirow{2}{*}{\textbf{$\tau$}} & \multirow{2}{*}{\textbf{\#Nodes}} & \multirow{2}{*}{\textbf{\#Edges}} & \multirow{2}{*}{\textbf{\#CC}} & \multirow{2}{*}{\textbf{\#Isolated Nodes}} & \multicolumn{4}{c}{\textbf{Degree}} & \multicolumn{4}{c}{\textbf{Size of Connected Component}} \\
        \cmidrule(lr){7-10} \cmidrule(lr){11-14}
        & & & & & & Min & Max & Mean & Median & Min & Max & Mean & Median \\
        \midrule
        10\% & 0.9252 & 100,000 & 26,879,058 & 64,417 & 58,010 & 0 & 5,872 & 268.79 & 0 & 1 & 12,054 & 1.55 & 1 \\
        15\% & 0.9436 & 100,000 & 17,964,868 & 71,235 & 65,126 & 0 & 4,803 & 179.65 & 0 & 1 & 10,703 & 1.40 & 1 \\
        20\% & 0.9555 & 100,000 & 6,197,746 & 78,359 & 73,300 & 0 & 4,527 & 123.95 & 0 & 1 & 7,960 & 1.28 & 1 \\
        \bottomrule
    \end{tabular}
    \end{center}
    
    \textsuperscript{1}The ratio of $\tau$ increased by \(\tau_0\); larger ratios maintains higher confidence of relationship but retain fewer edges.
\end{table*}

\begin{table*}[tb]
    \centering
    \caption{Summary of Node (Database) Properties in WikiDBGraph}
    \label{tab:node-props}
    \begin{tabular}{llccccl}
        \toprule
        \textbf{Category} & \textbf{Property} & \textbf{Min} & \textbf{Max} & \textbf{Mean} & \textbf{Median} & \textbf{Description} \\
        \midrule
        \multirow{5}{*}{Structural} & \#Tables & 2 & 100 & 16.11 & 3.0 & Number of tables in the database \\
        & \#Columns & 7 & 7,955 & 849.32 & 78.0 & sum of \#columns across all tables \\
        & CategoricalRatio & 0.4 & 1.0 & 0.92 & 0.92 & Proportion of categorical columns \\
        & ForeignKeyDensity & 0.0 & 0.36 & 0.05 & 0.04 & \#foreign keys / \#columns \\
        & AverageConnection & 1.0 & 13.5 & 2.71 & 1.50 & Average \#foreign keys per table \\
        \midrule
        \multirow{3}{*}{Semantic} & DatabaseEmbed & \multicolumn{5}{c}{Database embedding vector (768-dim)} \\
        & ClusterID & \multicolumn{5}{c}{Topic ID from clustering (12 clusters)} \\
        & CommunityID & \multicolumn{5}{c}{Community from graph structure (6133 communities)} \\
        \midrule
        \multirow{5}{*}{Statistical} & DataVolume & 12.3KB & 215.9MB & 1.5MB & 73.7KB & Total size of database files \\
        & AllJoinSize & 10.00 & $9\times10^{21}$ & $9\times10^{16}$ & 73.00 & Row count when joining all tables \\
        & AverageCardinality & 0.00 & $7.3\times10^4$ & 67.56 & 15.03 & Average distinct values per column \\
        & AverageSparsity & 0.00 & 0.58 & 0.20 & 0.21 & Average proportion of NULL values \\
        & AverageEntropy & 0.60 & 12.20 & 3.50 & 3.28 & Average entropy of columns \\
        \bottomrule
    \end{tabular}
\end{table*}

\begin{table*}[tb]
    \centering
    \caption{Summary of Edge (Database Relationship) Properties in WikiDBGraph ($\tau=0.94$)}
    \label{tab:edge-props}
    \begin{minipage}{\textwidth}
    \begin{center}
    \begin{tabular}{llccccl}
        \toprule
        \textbf{Category} & \textbf{Property} & \textbf{Min} & \textbf{Max} & \textbf{Mean} & \textbf{Median} & \textbf{Description} \\
        \midrule
        \multirow{8}{*}{Structural} & JaccardTable & 0.00 & 1.00 & 0.12 & 0.02 & Jaccard index of table name sets \\
        & JaccardColumn & 0.00 & 1.00 & 0.36 & 0.37 & Jaccard index of column name sets \\
        & JaccardType & 0.33 & 1.00 & 0.93 & 1.00 & Jaccard index of data type sets \\
        & HellingerDistanceType & 0.00 & 0.57 & 0.07 & 0.02 & Hellinger distance of data type distributions \\
        & GED & 0 & 1230 & 111.89 & 4 & Graph edit distance (GED) between schema graphs\textsuperscript{1} \\
        & CommonTables & 0 & 60 & 1.85 & 1 & \#common tables between schemas \\
        & CommonColumns & 0 & 1092 & 74.00 & 13 & \#common columns between schemas \\
        & CommonDataTypes & 2 & 6 & 3.24 & 2 & \#common data types between schemas \\
        \midrule
        \multirow{2}{*}{Semantic} & EmbedSim & 0.9436 & 0.9998 & 0.9732 & 0.9771 & Cosine similarity of database embeddings \\
& SimilarityConfidence & 0.9982 & 1.0000 & 0.9991 & 0.9991 & Confidence score\textsuperscript{2} of similarity prediction \\
        \midrule
        \multirow{2}{*}{Statistical} & KLDivergence & 0.00 & 34.01 & 17.05 & 17.60 & KL divergence of shared column distributions \\
& OverlapRatio & 0.00 & 1.00 & 0.22 & 0.20 & Ratio of overlapping values in shared columns \\
        \bottomrule
    \end{tabular}
\end{center}
    \textsuperscript{1}The schema graph is a directed graph that connects tables (nodes) with foreign keys (edges), which is distinct from WikiDBGraph.\\
    \textsuperscript{2}The ratio of edges which has lower similarities. \\
    \end{minipage}
\end{table*}

%% file: sections/experiment.tex
This section evaluates the utility of WikiDBGraph for collaborative learning. Section~\ref{subsec:exp_settings} details the experimental settings. Because existing CL algorithms~\cite{li2023fedtree,cheng2021secureboost} are not directly applicable end-to-end to WikiDBGraph, we design an automated data-mining pipeline (Section~\ref{subsec:auto-eval}) that interfaces with existing CL training algorithms. The performance on the most similar database pairs are presented in Section~\ref{subsec:auto-eval-perf}.

\subsection{Experimental Settings}\label{subsec:exp_settings}
This section describes the data, algorithms, licenses, and model/training hyperparameters used in our experiments.

\noindent\textbf{Data.}
We select the top 2{,}000 database pairs with the highest similarity scores from WikiDBGraph and, among them, choose two representative pairs that have distinct TIDs yet exhibit high similarity in schema embeddings.

\noindent\textbf{Algorithms.}
We evaluate representative collaborative learning algorithms spanning horizontal federated learning (HFL), vertical federated learning (VFL), and split learning. These algorithms include: \textit{Solo}: Training on data from a single database. \textit{FedProx}~\cite{li2020federated}: An HFL algorithm to mitigate data heterogeneity ($\mu=0.001$). \textit{FedOV}~\cite{diao2023towards}: An HFL method designed for data or feature overlap scenarios. \textit{SplitNN}~\cite{vepakomma2018split}: A split-learning algorithm for vertically partitioned data, where the model is split across clients. \rev{\textit{SFL}~\cite{chen2022personalized}: A personalized federated learning method on a graph of clients.} \textit{Combined}: Centralized training on all clients' data.

\noindent\textbf{Model and Training Hyperparameters.}
The base model for all algorithms is a two-layer neural network with [64, 32] hidden dimensions. For SplitNN, this architecture is partitioned such that each client's local model computes a 64-dimensional embedding, and a central server performs aggregation with the final 32-dimensional layer. All models are trained using the Adam optimizer~\cite{kingma2014adam} with a learning rate of $10^{-4}$ and a weight decay of $10^{-5}$.

\noindent\textbf{Evaluation Metrics.}
We partition the data into an 80\% training set and a 20\% testing set. To provide a comprehensive performance comparison, we report four standard classification metrics: Accuracy, Precision, Recall, and F1-score. All reported results are the mean values over five independent runs on the combined test sets from both clients to ensure robustness.

\noindent\textbf{Hardware.}
The experiments are conducted on a server equipped with AMD EPYC 9654 96-Core Processors (1.1\,TB system RAM) and four NVIDIA H100 PCIe GPUs (80\,GB HBM2e memory each).

\noindent\textbf{Licenses.}
The WikiDBGraph dataset is released under the Creative Commons Attribution 4.0 International (CC BY 4.0) license. The code of WikiDBGraph is released under the Apache License 2.0. 

\begin{revblock}

\subsection{Evaluation of Benchmark Design}\label{subsec:eval_benchmark}

This subsection evaluates the embedding-based relationship predictor against baselines and quantifies the contribution of each benchmark design choice via ablation studies.

\noindent\textbf{Effect of prompt design.} Table~\ref{tab:abl-data-schema} demonstrates the importance of prompting with both schema and data. The combined approach (Full) achieves the best overall performance, indicating that schema definitions and data samples offer complementary context.

\noindent\textbf{Effect of encoder model.} Table~\ref{tab:ablation-encoder-model} shows that the overall conclusions are consistent across encoder backbones and the string-based schema matching. Contrastive learning substantially improves AUC-ROC by 0.03--0.13 for different encoders. For instance, replacing BGE-M3 with all-mpnet-base-v2~\cite{song2020mpnet} yields comparable performance after contrastive learning, suggesting that the gains primarily come from the proposed contrastive objective rather than a specific pretrained encoder.

\noindent\textbf{Effect of negative samples.} Fig.~\ref{fig:abl_neg_sample} shows that varying the number of negative samples has only a marginal impact on AUC. In particular, performance does not improve further beyond the default setting and slightly degrades when substantially increasing the number of negatives. We attribute this to the class imbalance in our data, where overly many negatives may bias the contrastive objective toward easy non-matches.

\noindent\textbf{Effect of sample size in prompt.} Fig.~\ref{fig:abl_sample_size} shows that, once at least one tuple is included, the number of sampled tuples in the training prompt has only a limited impact on performance. Increasing the sample size primarily lengthens the context and raises computation cost, while yielding negligible improvements in AUC.

\noindent\textbf{Effect of similarity threshold.} Fig.~\ref{fig:threshold_graph} analyzes how the similarity threshold $\tau$ affects graph construction. As $\tau$ decreases, both the average node degree (Fig.~\ref{fig:abl_threshold_degree}) and the total number of edges (Fig.~\ref{fig:abl_threshold_edge}) grow rapidly, shifting the graph from sparse (dominated by highly similar pairs) to dense (containing many mid-similarity edges). Since most existing CL methods primarily work on highly similar pairs, we adopt a relatively high $\tau$ in our main setting for methodological compatibility. When more advanced CL techniques become available for mid-similarity regimes, our benchmark can readily support them by evaluating on pairs induced by lower thresholds.

\begin{table}[ht]
    \centering
    \caption{Performance of relationship prediction  (optimal threshold, 5 seeds). Contrast.: contrastive learning. Full: Schema + Data}
    \label{tab:abl-data-schema}
    \resizebox{\columnwidth}{!}{
    \setlength{\tabcolsep}{3pt}
    \begin{tabular}{llccc}
        \toprule
        \textbf{Configuration} & \textbf{Method} & \textbf{Accuracy} & \textbf{AUC-ROC} & \textbf{F1} \\
        \midrule
        BGE-M3 & Original & 0.9016±0.0026 & 0.9663±0.0012 & 0.9005±0.0029 \\
        \midrule
        Schema Only & Contrast. & 0.9622±0.0015 & 0.9908±0.0006 & 0.9618±0.0014 \\
        Data Only & Contrast. & \underline{0.9766±0.0011} & \underline{0.9964±0.0004} & \underline{0.9765±0.0012} \\
        \midrule
        BGE-M3 + Full & Contrast. & \textbf{0.9772±0.0007} & \textbf{0.9967±0.0004} & \textbf{0.9770±0.0007} \\
        \bottomrule
    \end{tabular}}
\end{table}

\begin{table}[ht]
    \centering
    \caption{Different encoder model's test performance (optimal threshold, 5 seeds). Contrast.: contrastive learning.}
    \label{tab:ablation-encoder-model}
    \resizebox{\columnwidth}{!}{
    \setlength{\tabcolsep}{3pt}
    \begin{tabular}{llccc}
        \toprule
        \textbf{Encoder Model} & \textbf{Method} & \textbf{AUC-ROC} & \textbf{Precision} & \textbf{Recall} \\
        \midrule
        String Match (Jaccard) & - & 0.7367±0.0051 & 0.7345±0.0117 & 0.5661±0.0025 \\
        \midrule
        \multirow{2}{*}{BGE-M3} & Original & 0.9663±0.0012 & 0.9107±0.0037 & 0.8905±0.0066 \\
         & Contrast. & \underline{0.9967±0.0004} & \textbf{0.9851±0.0029} & \underline{0.9689±0.0027} \\
        \midrule
        \multirow{2}{*}{all-mpnet-base-v2} & Original & 0.8667±0.0027 & 0.8369±0.0050 & 0.7401±0.0044 \\
         & Contrast. & \textbf{0.9968±0.0004} & \underline{0.9846±0.0033} & \textbf{0.9827±0.0044} \\
                \bottomrule
    \end{tabular}}
\end{table}

\begin{figure}[ht]
    \centering
     \begin{subfigure}[b]{0.48\linewidth}
         \centering
         \includegraphics[width=\textwidth]{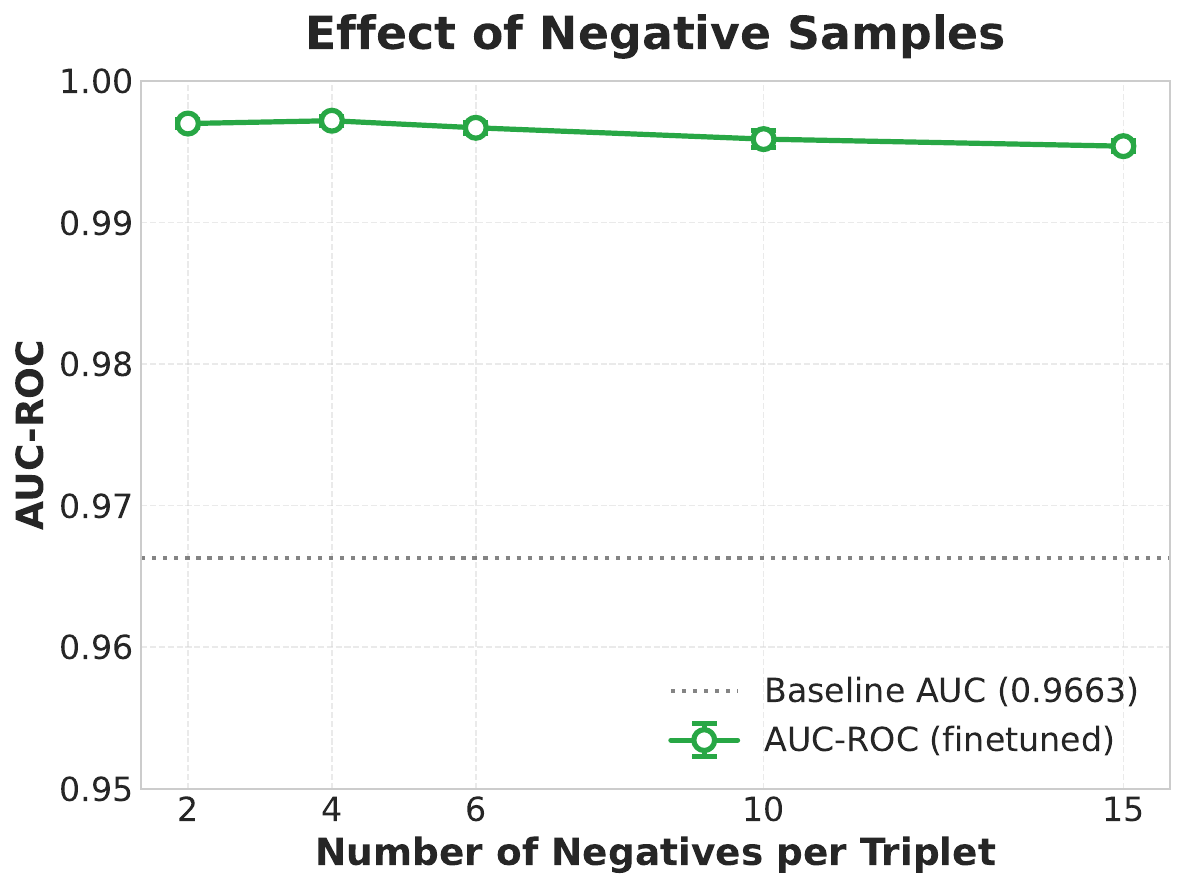}
         \caption{\#Negative samples}
         \label{fig:abl_neg_sample}
     \end{subfigure}
     \hfill %
     \begin{subfigure}[b]{0.48\linewidth}
         \centering
         \includegraphics[width=\textwidth]{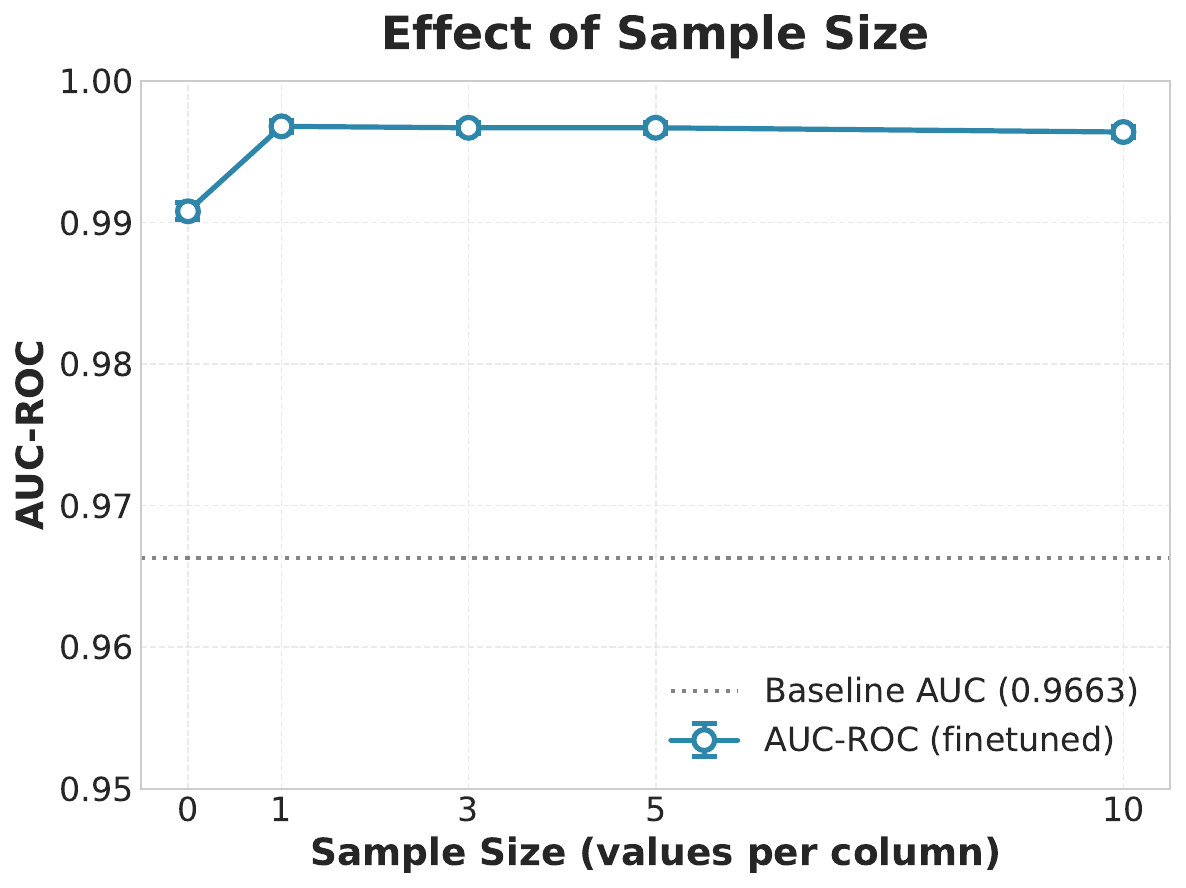}
         \caption{\#Sample size}
         \label{fig:abl_sample_size}
     \end{subfigure}
     
     \caption{Effect of hyperparameters on predictive AUC}
    \label{fig:hyperparam_auc}
\end{figure}

\begin{figure}[ht]
    \centering
     \begin{subfigure}[b]{0.48\linewidth}
         \centering
         \includegraphics[width=\textwidth]{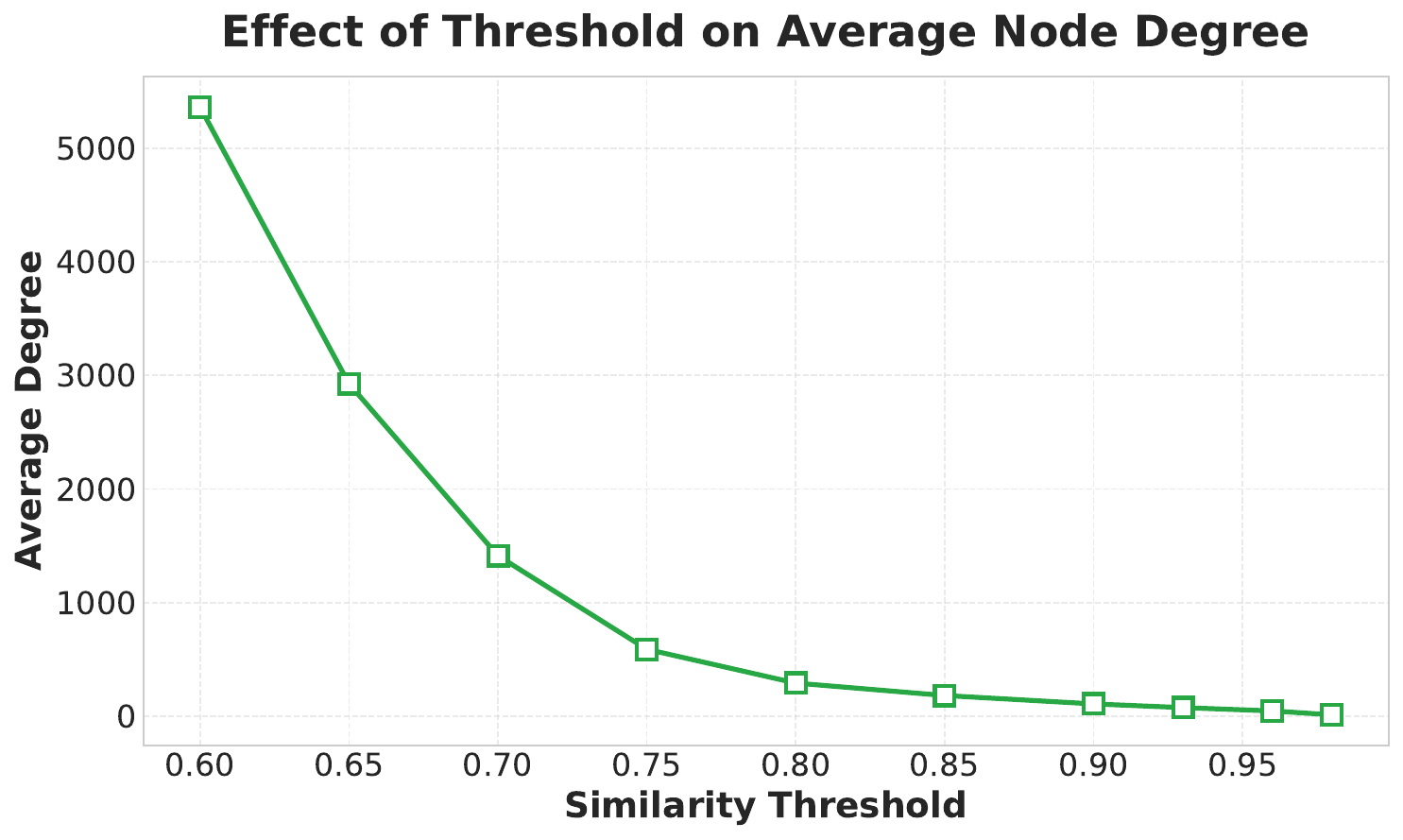}
         \caption{Averaged degree}
         \label{fig:abl_threshold_degree}
     \end{subfigure}
     \hfill %
     \begin{subfigure}[b]{0.48\linewidth}
         \centering
         \includegraphics[width=\textwidth]{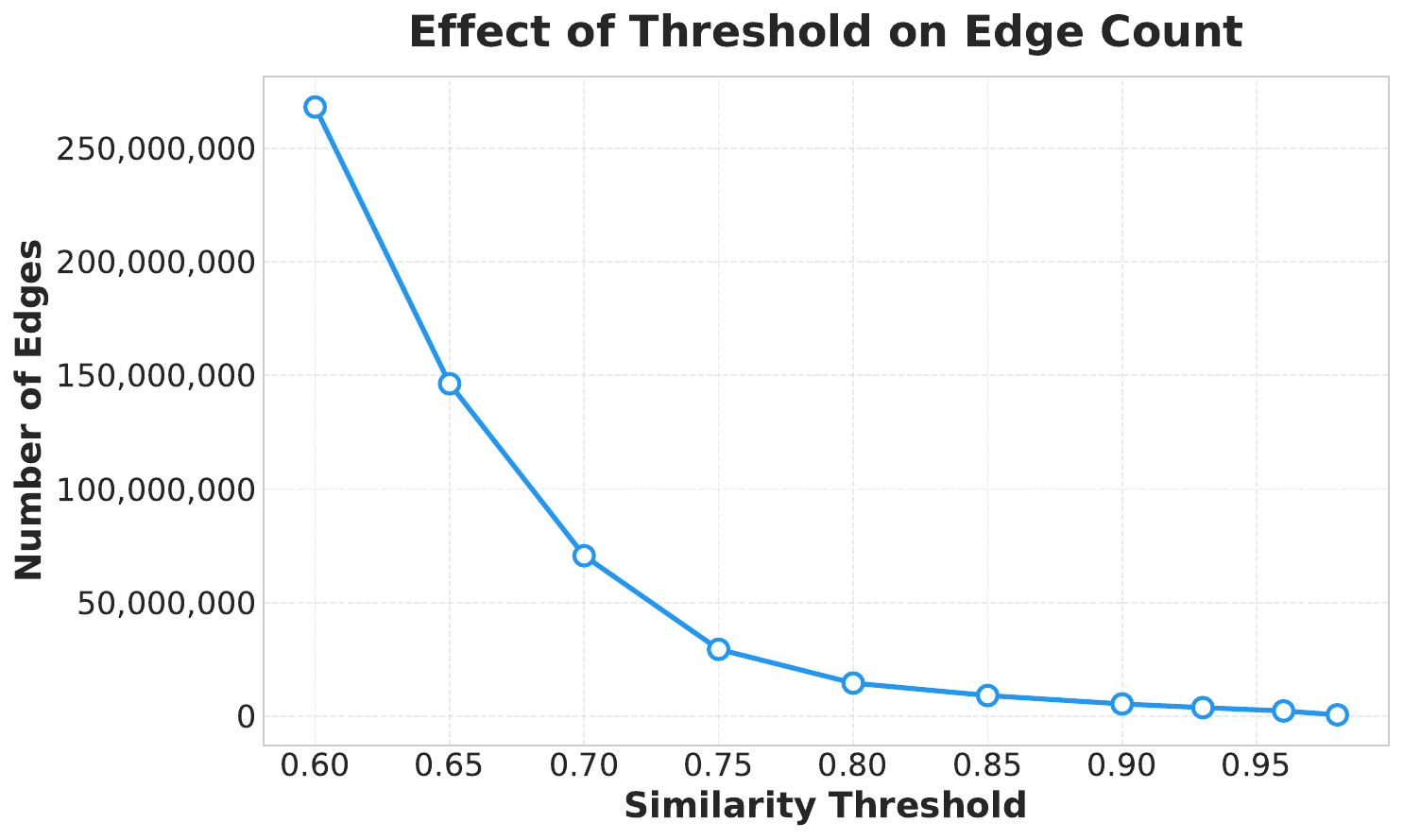}
         \caption{Total \#edges}
         \label{fig:abl_threshold_edge}
     \end{subfigure}
     
     \caption{Effect of threshold $\tau$ on the graph properties}
    \label{fig:threshold_graph}
\end{figure}

\end{revblock}

\subsection{Automated Data Mining and Evaluation}\label{subsec:auto-eval}

We employ an automated data mining and evaluation pipeline to validate whether CL on database pairs identified by WikiDBGraph yields a performance improvement over \textit{Solo}. This pipeline consists of five sequential stages. The detailed algorithm is presented in Algorithm~\ref{alg:auto-eval}.

The first stage, \textbf{database pair sampling}, selects the 2,000 pairs with the highest embedding similarity scores from all candidates that satisfy our validation criteria---both databases contain at least one table with a minimum of 100 rows. The second stage, \textbf{label selection}, identifies a target column for the learning task. We prioritize classification for simplicity; among all columns with a unique value count between 2 and 50, we select the one with the least number of categories to serve as the label. The third stage, \textbf{table joining}, constructs a comprehensive table for each database by left-joining its tables according to their foreign key relationships, starting from the table containing the selected label. To manage memory, this join process is halted if the table's instance count exceeds one million. These heuristics enable a practical ``zero-human-in-the-loop'' pipeline, where the 2–50 value constraint filters out identifiers and continuous variables to ensure uniform classification tasks.

We then remove constant columns and any column with more than 50\% missing values, followed by feature standardization. The fourth stage is \textbf{column alignment}, where we harmonize the schemas of the two resulting tables by retaining only their common columns. Column names are matched in a case-insensitive manner after ignoring all non-alphanumeric symbols. \rev{As an optional final quality-control step, we manually inspect each candidate task for semantic relatedness, column-alignment correctness, and label reasonableness. In total, 1,420 of 2,000 tasks pass and are used for training and evaluation. This step does not affect our conclusions, because the rejected tasks are dominated by invalid or non-executable cases already invalidated by the automatic pipeline (e.g., near-constant labels, too few non-constant features after joining/filtering, or runtime failures that prevent CL from training).} Finally, the last stage is \textbf{training and evaluation}. For each client, the aligned data are partitioned into an 80\%/20\% train/test split. All CL algorithms are trained on the clients' aligned training sets. To assess improvement, every algorithm---including \textit{Solo} and \textit{Combined}---is evaluated on the combined test set from all the clients.

\begin{revblock}
In later stages, column alignment can be tackled via schema matching methods (e.g., \cite{khatiwada2023santos,fan2022semantics}) that explicitly recover semantic correspondences between columns. While \cite{fan2022semantics} also employs a contrastive learning framework, it focuses on generating fine-grained contextualized column embeddings to resolve feature overlap, distinct from our work learning whole-database embeddings for global structural representation. Additionally, TURL \cite{deng2022turl} provides semantic table representations that facilitate table augmentation, further benefiting vertical alignment in instance-overlapped settings. Integrating these schema-level signals into future automated CL systems can significantly enhance overall integration performance.
\end{revblock}

\begin{algorithm}[htbp]
    \SetAlgoLined
    \small
    \caption{Automated Evaluation Pipeline}\label{alg:auto-eval}
    \KwIn{Database pairs with similarity scores; algorithm set $\mathcal{A}\subseteq\{\textit{FedAvg},\textit{FedProx},\textit{FedOV},\textit{SplitNN},\textit{Combined}\}$}
    \KwOut{Trained models and performance of \textit{Solo} and all $a\in\mathcal{A}$ on a combined test set}
    
    \tcc{Stage 1: Database-pair sampling}
    $\mathcal{P} \leftarrow$ select top $2{,}000$ database pairs with $\ge100$ rows\;
    $\mathcal{R} \leftarrow \emptyset$\;

    \ForEach{pair $(D_A, D_B) \in \mathcal{P}$}{
        \tcc{Stage 2--4: clean and align}
        \ForEach{$D \in \{D_A, D_B\}$}{
            $y_D \leftarrow \textbf{SelectLabel}(D)$\;
            $T_D^{\mathrm{clean}} \leftarrow \textbf{JoinTables}(D, y_D)$\tcp*{halt at 1M rows}
        }
        $(X_A', X_B') \leftarrow \textbf{AlignColumns}(T_A^{\mathrm{clean}}, T_B^{\mathrm{clean}})$\;
        $X_A^{\mathrm{train}}, X_A^{\mathrm{test}}, X_B^{\mathrm{train}}, X_B^{\mathrm{test}} \leftarrow$ partition $X_A', X_B'$ by $4{:}1$\;
        $X^{\mathrm{test}} \leftarrow X_A^{\mathrm{test}} \cup X_B^{\mathrm{test}}$\;

        \tcc{Stage 5: train and evaluate}
        $M^{\mathrm{Solo}} \leftarrow \textbf{Train}(\textit{Solo} \text{ on } X_A^{\mathrm{train}})$\;
        $\mathrm{Perf}^{\mathrm{Solo}} \leftarrow \textbf{Evaluate}(M^{\mathrm{Solo}} \text{ on } X^{\mathrm{test}})$\;
        \ForEach{$a \in \mathcal{A}$}{
            $M[a] \leftarrow \textbf{Train}\big(a \text{ on aligned } (X_A^{\mathrm{train}}, X_B^{\mathrm{train}})\big)$\;
            $\mathrm{Perf}[a] \leftarrow \textbf{Evaluate}(M[a] \text{ on } X^{\mathrm{test}})$\;
        }
        $\mathcal{R}[(D_A,D_B)] \leftarrow \{\mathrm{Perf}^{\mathrm{Solo}}, \mathrm{Perf}[\cdot]\}$\;
    }

    \Return $\mathcal{R}$\;
\end{algorithm}

\subsection{Performance of Existing CL Algorithms}\label{subsec:auto-eval-perf}

The performance distributions of different CL algorithms are shown in Fig.~\ref{fig:fl_performance_scores}. We report only the performance for the 1{,}207 that successfully completed training. Two primary observations emerge. First, the average scores of CL algorithms generally lie between the \textit{Solo} and \textit{Combined} baselines. Second, performance differences among CL algorithms are small; the notable exception is \textit{SCAFFOLD}, which is known to be sensitive to hyperparameters and was not carefully tuned here. Both observations align with findings reported in NIIDBench~\cite{li2022federated}, a well-known federated learning benchmark, indicating that evaluation using WikiDBGraph is effective and consistent with prior studies.

Despite the improvement in average performance, the generalization of existing CL algorithms to practical tasks---even when coupled with the automated data-mining pipeline (Alg.~\ref{alg:auto-eval})---remains limited. Table~\ref{tab:fl_improvement_summary} summarizes performance improvements across the 1{,}207 successful tasks: fewer than half of the tasks benefit from CL.

The high rate of non-improvement indicates that the primary bottleneck lies in data preprocessing. Our automated pipeline---which uses simple string-based column alignment---can miss semantic matches or, worse, join dissimilar columns. This flawed preparation yields a ``garbage in, garbage out'' effect: training on misaligned data degrades collaborative performance below the single-client baseline. Section~\ref{sec:case_study} supports this conclusion: manually curated alignments on similar database pairs produce substantial gains, implicating the pipeline rather than the CL algorithms.

While severe data heterogeneity (non-IID)---a known challenge for standard algorithms like FedAvg---is certainly a contributing factor, the stark performance gap between these automated results and the significant gains seen in our manually-curated case studies (Tables~\ref{tab:horizontal_fl_comparison}, \ref{tab:vertical_fl_comparison}) points to preprocessing as the more immediate and critical challenge.

Two major limitations of the current preprocessing pipeline are table joining and column alignment. First, the current implementation of the table-joining step is capped at one million rows to manage memory, which can discard valuable data; addressing this likely requires redesigning novel CL algorithms that take databases, rather than prejoined tables, as input. Second, the column-alignment step by comparing string similarity is straightforward but may be suboptimal because it ignores semantic similarity; addressing this calls for a new efficient, semantics-aware column-alignment method to determine which columns are equivalent. \rev{To validate this, we conduct an additional experiment using DeepJoin~\cite{dong2023deepjoin} for semantics-aware column alignment, while still using BGE-M3 as the base embedding model. As shown in Figure~\ref{fig:raw_vs_semantic}, on the same task set (with identical labels), semantic-aware alignment improves the mean F1 score by 0.06--0.09 for FedProx, FedOV, and FedAvg.}\looseness-1

\begin{revblock}

\begin{figure}
    \centering
    \includegraphics[width=0.98\linewidth]{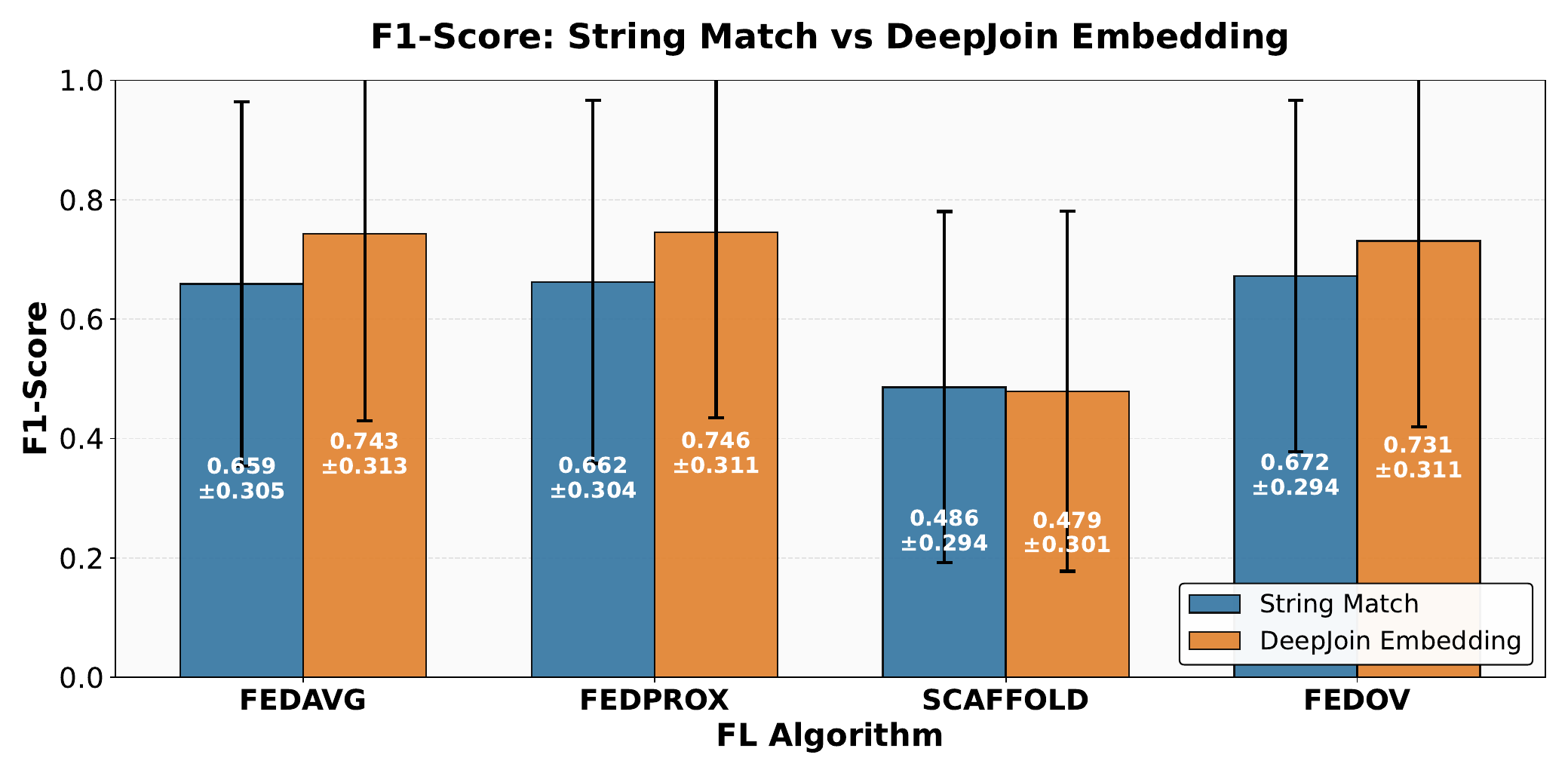}
    \caption{F1 improvement of DeepJoin for column alignment}
    \label{fig:raw_vs_semantic}
\end{figure}

\end{revblock}

\begin{revblock}

\noindent\textbf{Quantification of Non-IID Heterogeneity.} To rigorously compare against synthetic non-IID benchmarks~\cite{diao2023towards}, we model the label skew as a multivariate Dirichlet process. We estimate the concentration parameter $\alpha$ using the robust Method of Moments~\cite{minka2000estimating}, which aligns the theoretical Dirichlet variance with the empirical variance observed across WikiDBGraph pairs. The resulting distribution (Fig.~\ref{fig:dist_dirichlet_alpha}) reveals a long-tailed spectrum of heterogeneity. While a substantial portion of pairs exhibit near-IID characteristics ($\alpha > 5.0$), we identify a critical cluster of highly heterogeneous data: $24.7\%$ of pairs fall into the high-heterogeneity regime ($\alpha < 1.0$), with $18.4\%$ exhibiting extreme non-IID skew ($\alpha < 0.5$). This confirms that WikiDBGraph does not merely replicate standard distributions but captures the sharp, ``long-tail'' non-IID challenges inherent in real-world data silos.

\begin{figure}[ht]
    \centering
    \includegraphics[width=0.98\linewidth]{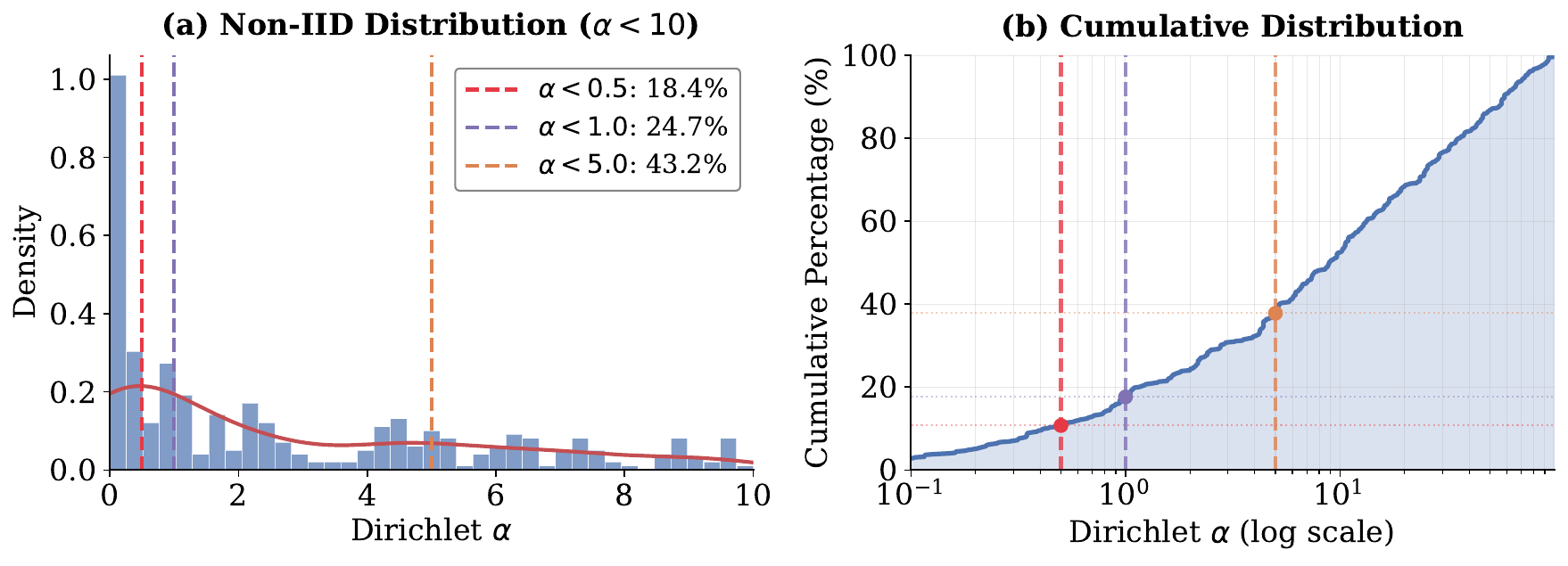}
    \caption{Distribution of estimated Dirichlet-$\alpha$ in WikiDBGraph}
    \label{fig:dist_dirichlet_alpha}
\end{figure}

\end{revblock}

\begin{revblock}
\begin{figure*}[tb]
    \centering
    \begin{subfigure}[b]{0.49\linewidth}
        \centering
        \includegraphics[width=\linewidth]{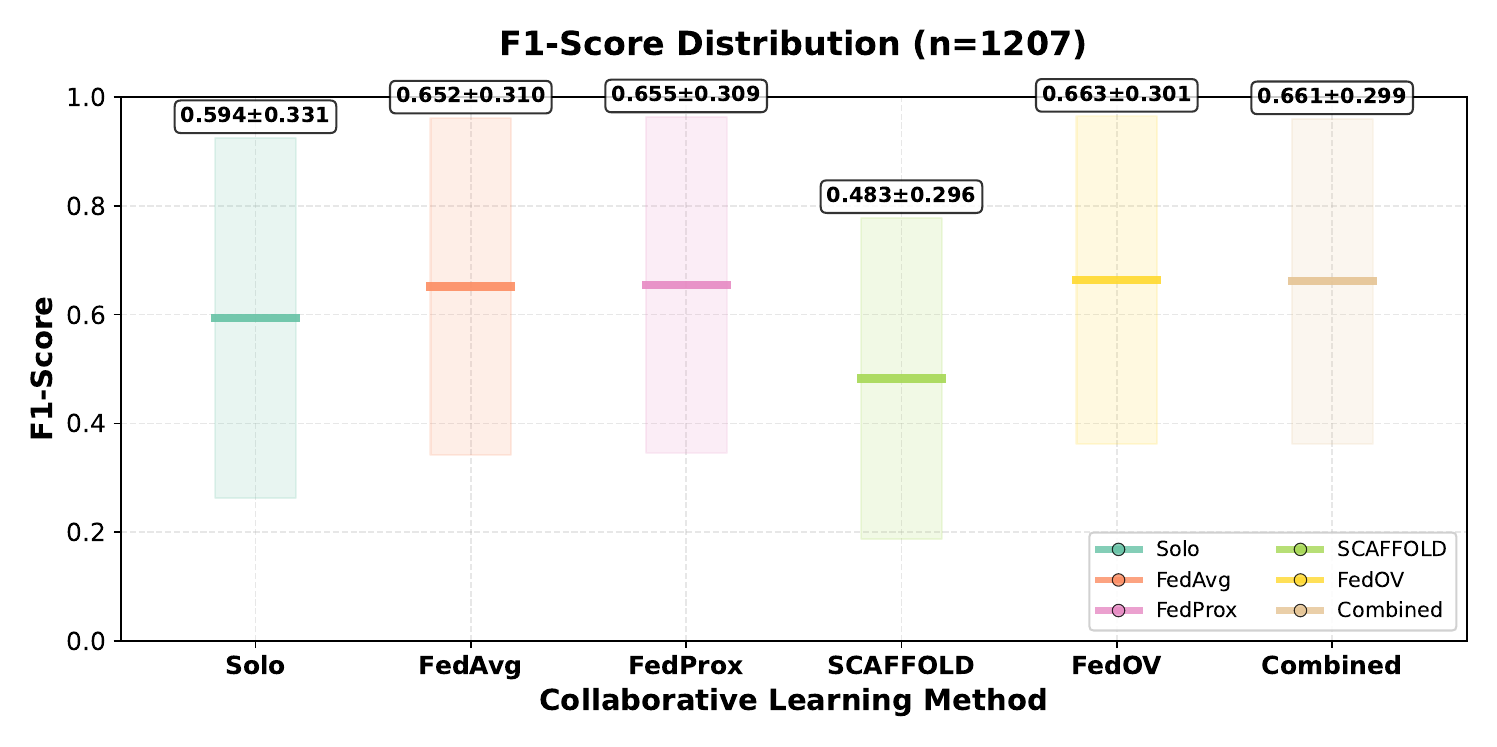}
        \caption{Distribution of F1-Score.}
        \label{fig:final_f1_distribution}
    \end{subfigure}
    \hfill
    \begin{subfigure}[b]{0.49\linewidth}
        \centering
        \includegraphics[width=\linewidth]{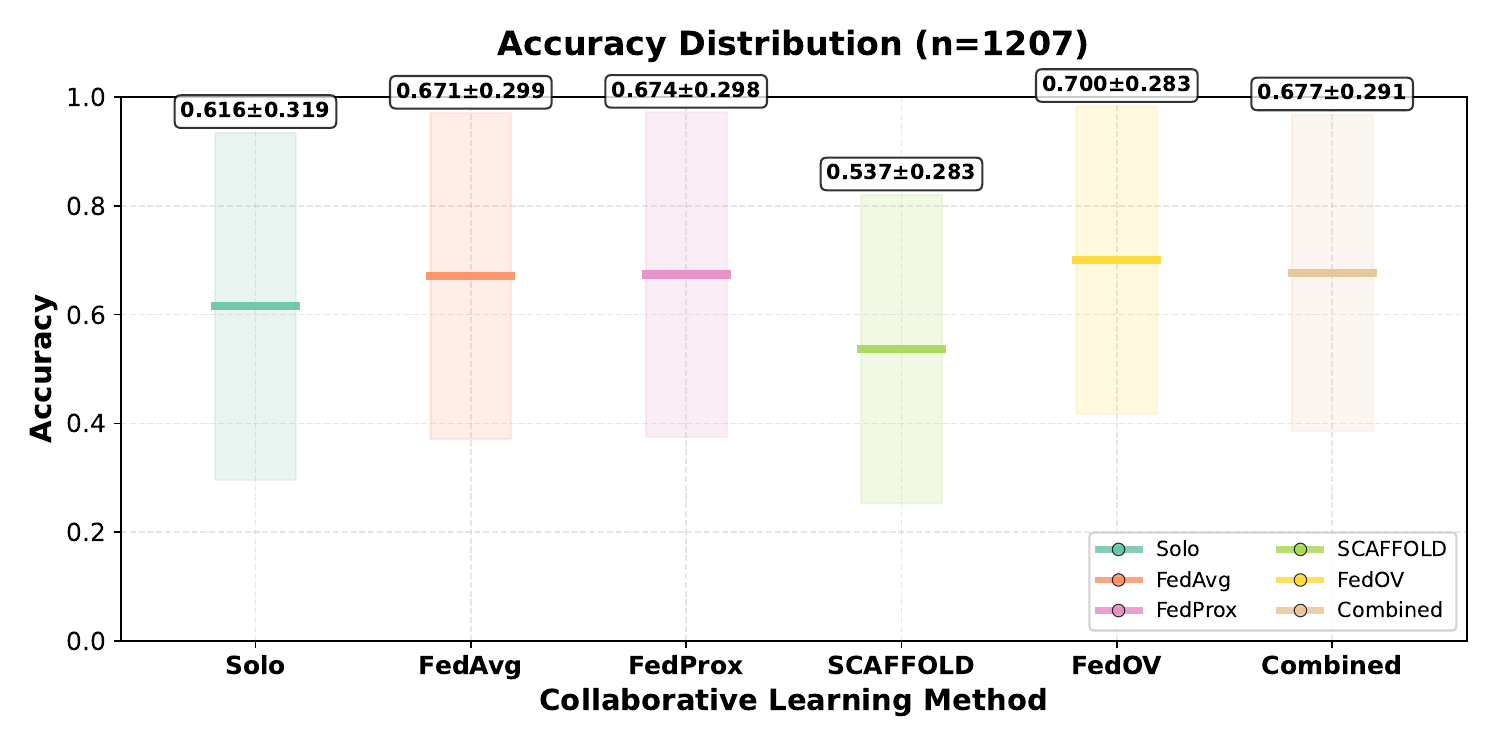}
        \caption{Distribution of Accuracy.}
        \label{fig:final_accuracy_distribution}
    \end{subfigure}
    \caption{Final metric distributions achieved by \textit{Solo}, \textit{Combined}, and different CL algorithms}
    \label{fig:fl_performance_scores}
\end{figure*}
\end{revblock}

\begin{table}[tb]
    \centering
    \caption{Summary of automated data mining results with different CL algorithms on 2,000 database pairs}
    \label{tab:fl_improvement_summary}
    \begin{tabular}{lccc}
    \toprule
    \textbf{Metric} & \textbf{Improved vs. Solo} & \textbf{Total} & \textbf{Ratio} \\
    \midrule
    FedAvg $>$ Solo & 566 & 1207 & 46.9\% \\
    FedProx $>$ Solo & 576 & 1207 & 47.7\% \\
    SCAFFOLD $>$ Solo & 405 & 1207 & 33.6\% \\
    FedOV $>$ Solo & 559 & 1207 & 46.3\% \\
    Combined $>$ Solo & 565 & 1207 & 46.8\% \\
    \bottomrule
    \end{tabular}
\end{table}

%% file: sections/case_study.tex
This section includes three case studies using manually curated data: \textit{feature overlap} (Section~\ref{subsec:case1}), \textit{instance overlap} (Section~\ref{subsec:case2}), and hybrid overlap (Section~\ref{subsec:case-study-hybrid}). All cases are novel findings enabled by WikiDBGraph, as their corresponding database pairs have distinct TIDs and no explicit links in Wikidata or WikiDBs.

\subsection{Case 1: Two Databases with Feature Overlap}\label{subsec:case1}

We exemplify the feature overlap scenario with a database pair: 02799 (\textit{Trypanosoma\_Cruzi\_Orthologs\_Db\_30}, TID: Q62194121) and 79665 (\textit{TrypanosomaCruziOrthologs225}, TID: Q62256692). This pair exhibits a maximum embedding similarity score (EmbedSim) of 0.9894. Both databases possess nearly identical schemas, each containing two tables with the same 24 columns and corresponding foreign key relationships, which constitutes the feature overlap. The primary distinction lies in their data instances and distributions; database 02799 contains 282 rows in both its tables, whereas database {79665} has 514 and 304 rows, respectively. This disparity makes the pair an ideal candidate for studying collaborative learning such as HFL. We therefore evaluate on a multi-class classification task to predict the label---\textit{details\_encoded\_protein} with six distinct classes.

We first evaluate the performance of several neural network (NN) based HFL algorithms, with results summarized in Table~\ref{tab:horizontal_fl_comparison}. From these results, we draw two primary conclusions. First, \textbf{all HFL methods consistently and significantly outperform at least one of the \textit{Solo} baselines}. This confirms the effectiveness of WikiDBGraph in finding new databases that enable effective collaborative training. Second, \textbf{a notable performance gap persists between all HFL algorithms and the centralized training baseline (\textit{Combined})}. This behavior is characteristic of the challenges posed by non-identically and independently distributed (non-i.i.d.) data, a well-known issue in federated learning~\cite{li2022federated}.

To demonstrate broader applicability, we extend our evaluation to tree-based models, known as \textit{FedTree}~\cite{li2023fedtree} (Table~\ref{tab:fedtree_horizontal_fl}). The results consistently show that both the \textit{Combined} and \textit{FedTree} models outperform their respective \textit{Solo} baselines. This indicates that the benefits of collaboration extend beyond NN, further validating the effectiveness of WikiDBGraph in identifying valuable opportunities for collaborative learning.

\begin{table}[tb]
    \centering
    \caption{Horizontal federated learning accuracy on feature-overlapped databases (six-class classification)}
    \label{tab:horizontal_fl_comparison}
    \begin{tabular}{lcccc}
    \toprule
    \textbf{Method} & \textbf{Accuracy} & \textbf{Precision} & \textbf{Recall} & \textbf{F1-score} \\
    \midrule
  Solo (DB 02799) & 0.2910 & 0.0980 & 0.2910 & 0.1443 \\
  Solo (DB 79665) & \underline{0.6187} & 0.4160 & \underline{0.6187} & 0.4961 \\
  \midrule
  FedAvg & {0.6154} & 0.6045 & {0.6154} & 0.5986 \\
  FedProx & {0.6154} & \underline{0.6045} & {0.6154} & \underline{0.5986} \\
  FedOV & 0.5753 & 0.5048 & 0.5753 & 0.5377 \\
    \midrule
    Combined & \textbf{0.8997} & \textbf{0.8130} & \textbf{0.8997} & \textbf{0.8532} \\
    \bottomrule
    \end{tabular}
  \end{table}

\begin{table}[tb]
  \centering
  \caption{FedTree accuracy on feature overlap}
  \label{tab:fedtree_horizontal_fl}
  \begin{tabular}{lcccc}
  \toprule
  \textbf{Method} & \textbf{Accuracy} & \textbf{Precision} & \textbf{Recall} & \textbf{F1-Score} \\  \midrule
  Solo (XGBoost DB1) & 0.1720 & 0.1720 & 0.1720 & 0.1720 \\
  Solo (XGBoost DB2) & 0.5071 & 0.5071 & 0.5071 & 0.5071 \\
  \midrule
  FedTree~\cite{li2023fedtree} & \underline{0.8428} & \underline{0.8176} & \underline{0.8428} & \underline{0.8172} \\
  \midrule
  Combined (XGBoost) & \textbf{0.9674} & \textbf{0.9674} & \textbf{0.9674} & \textbf{0.9674} \\
  \bottomrule
  \end{tabular}
\end{table}

\subsection{Case 2: Two Databases with Instance Overlap}
\label{subsec:case2}

We demonstrate an instance overlap scenario with two databases of gene: {00381} (\textit{TrypanosomaCruziOrthologs1}, TID: Q62194121) and {48804} (\textit{Ortholog\_Lpg1l\_Genomic\_Data}, TID: Q62256692), which exhibit a high embedding similarity (EmbedSim) of 0.95. Our analysis reveals that two tables, \textit{GeneOrthologsAnnotations} from DB {00381} (the gene table) and \textit{Ortholog\_Lpg1l\_Protein\_Annotations} from DB {48804} (the protein table), share a significant number of instances based on their common \textit{GeneId} column. We designate the protein table as the primary table for our task and the gene table as a secondary, feature-enriching table. We conduct a classification task that predicts the \textit{lpg\_Uni\_Prot\_Protein\_Id} column from the protein table, which has 49 distinct categories.

\begin{table}[tb]
    \centering
    \caption{SplitNN accuracy on instance-overlapped databases (49-class classification)}
    \label{tab:vertical_fl_comparison}
    \begin{tabular}{lcccc}
    \toprule
    \textbf{Method} & \textbf{Accuracy} & \textbf{Precision} & \textbf{Recall} & \textbf{F1-score} \\
    \midrule
    Solo (DB 48804) & 0.4082 & 0.3062 & 0.4082 & 0.3374 \\
    \midrule
    SplitNN & \underline{0.6531} & \underline{0.5727} & \underline{0.6531} & \underline{0.5962} \\
    \midrule
    Combined & \textbf{0.6782} & \textbf{0.6437} & \textbf{0.6782} & \textbf{0.6587} \\
    \bottomrule
    \end{tabular}
\end{table}

The results of NN-based algorithms, summarized in Table~\ref{tab:vertical_fl_comparison}, yield two key findings. First, the \textbf{split learning approach significantly outperforms the \textit{Solo} baseline} across all evaluation metrics. This result validates our hypothesis that enriching the feature set by incorporating data from the correlated gene table leads to substantial performance gains. Second, we observe \textbf{a significant performance gap between \textit{SplitNN} and Combined}---the centralized baseline that represents the performance upper bound. While \textit{SplitNN} captures the majority of the potential improvement, this gap highlights the inherent challenges of distributed model training compared to a centralized approach. In summary, these findings strongly underscore the practical value of leveraging instance-overlapped databases for collaborative learning evaluation.

Further experiments with a tree-based VFL algorithm, \textit{SecureBoost}~\cite{cheng2021secureboost}, are presented in Table~\ref{tab:secureboost_vertical_fl}, reinforcing our primary findings: across different algorithms, including specialized ones, extending the feature set consistently yields a significant performance boost. This demonstrates the general applicability of {WikiDBGraph} and further validates its effectiveness in identifying valuable pairs for feature enrichment.

\begin{table}[tb]
  \centering
  \setlength{\tabcolsep}{2.5pt}
  \caption{SecureBoost accuracy on instance overlap}
  \label{tab:secureboost_vertical_fl}
  \begin{tabular}{lcccc}
  \toprule
  \textbf{Method} & \textbf{Accuracy} & \textbf{Precision} & \textbf{Recall} & \textbf{F1-Score} \\
  \midrule
  Solo & \underline{0.8418} & \underline{0.8418} & \underline{0.8418} & \underline{0.8418} \\
  {SecureBoost~\cite{cheng2021secureboost} (Combined)} & \textbf{0.9252} & \textbf{0.9252} & \textbf{0.9252} & \textbf{0.9252} \\
  \bottomrule
  \end{tabular}
\end{table}

\begin{revblock}

\subsection{Case 3: Multiple Databases with Instance Overlap}\label{subsec:case-study-multi-instance}

We consider a personalized CL classification task that reflects a policy analytics application. Each client corresponds to an independent country-profile database (DB10626, DB13841, DB17587, DB57240, DB73136) where features overlap only partially. The shared objective is to predict the minimum compulsory education age from locally stored attributes. To ensure the target is well-defined and comparable across databases, we discretize numeric ages into four ordinal classes (<5, 5, 6, 7). These five databases form a five-node, sixteen-edge subgraph, equipped with the WikiDBGraph node and edge properties. We then train a GNN-based CL method (SFL~\cite{chen2022personalized}) that consumes these graph properties; the results are reported in Table~\ref{tab:fedgnn-results}.

\begin{table}[ht]
    \centering
    \caption{SFL experiment results (per-DB averages over 5 seeds). Best in \textbf{bold}, second-best \underline{underlined}. Gain (\%) denotes the percentage of DBs where the method outperforms Solo, averaged over seeds.}
    \label{tab:fedgnn-results}
    \resizebox{\columnwidth}{!}{
    \setlength{\tabcolsep}{4pt}
    \begin{tabular}{lcccc}
        \toprule
        & \multicolumn{2}{c}{\textbf{Accuracy}} & \multicolumn{2}{c}{\textbf{F1}} \\
        \cmidrule(lr){2-3} \cmidrule(lr){4-5}
        \textbf{Method} & Mean$\pm$Std & Gain(\%) & Mean$\pm$Std & Gain(\%) \\
        \midrule
        Solo (No Federation) & 0.8930±0.0636 & N/A & 0.8708±0.0671 & N/A \\
        FedAvg & 0.9542±0.0444 & 20.0\% & 0.9373±0.0581 & 20.0\% \\
        SFL (No Properties) & 0.9358±0.0509 & 16.0\% & 0.9138±0.0691 & 16.0\% \\
        SFL (Node Only) & 0.9145±0.0756 & 16.0\% & 0.8979±0.0838 & 16.0\% \\
        SFL (Edge Only) & \underline{0.9559±0.0457} & \textbf{24.0\%} & \underline{0.9407±0.0606} & \textbf{24.0\%} \\
        SFL (Both) & \textbf{0.9559±0.0498} & \underline{24.0\%} & \textbf{0.9410±0.0655} & \underline{24.0\%} \\
        \bottomrule
    \end{tabular}}
\end{table}

Table~\ref{tab:fedgnn-results} supports two key findings. \textbf{(1) Node and edge properties are critical for graph-based CL, enabling gains beyond FedAvg}: enriching SFL~\cite{chen2022personalized} with node and/or edge attributes consistently improves both Accuracy and F1 over the no-property variant and yields the best overall performance, outperforming FedAvg. Notably, edge properties contribute the largest gains—the edge-only configuration nearly matches the full model—suggesting that pairwise cross-database signals are particularly informative for guiding message passing and personalization. \textbf{(2) Graph-based CL improves over Solo, but structure alone is insufficient}: SFL without any graph properties already improves over Solo, indicating that leveraging client connectivity is beneficial; however, it remains slightly below FedAvg, implying that the client graph structure by itself is not enough to beat federated learning baselines without informative node/edge properties.

\end{revblock}

\subsection{Case 4: Multiple databases with Hybrid Overlap}\label{subsec:case-study-hybrid}

To explore complex hybrid-overlap scenarios, we randomly select a small connected component (size 3 to 10) from WikiDBGraph for visualization (Fig.~\ref{fig:collab}). The subgraph links a national monuments database (15418) to a cluster of historic-places registries (54379, 37176, 85770, 50469). Within the registry cluster, the databases are largely \textit{horizontally} aligned, sharing many features such as \texttt{AdministrativeEntity}, \texttt{AreaSize}, and \texttt{EstablishmentDate}. The monuments database (15418) is further enriched by the national-treasure heritage registry (78243); these two databases share 542 columns (e.g., \texttt{CategoryLabel}, \texttt{ArchitecturalStyle}), indicating strong horizontal alignment. In contrast, 78243 and 85770 share only 32 columns---including identifiable attributes such as \texttt{Latitude}, \texttt{Longitude}, and \texttt{AdministrativeEntity}---and their instances do not fully overlap, yielding a \textit{partially vertical} connection with limited instance overlap. The pair (78243, 85770) thus bridges two horizontally aligned clusters (national treasures and historic places), enabling collaborative tasks such as predicting the national-treasure category for historic places.

This component exemplifies a \textbf{hybrid} alignment pattern---simultaneous horizontal alignment within clusters and partial vertical alignment across clusters with incomplete instance overlap---that is not supported by existing CL algorithms. Consequently, graph-aware methods that operate over partially aligned, multi-database structures remain an open problem.

%% file: sections/conclusion.tex
This paper presented WikiDBGraph, a large-scale graph of 100,000 interconnected databases designed to address the critical lack of inter-database relationships in existing corpora. We demonstrated that WikiDBGraph successfully identifies correlated databases that benefit from collaborative learning, while also highlighting the performance gap compared to centralized models and surfacing novel challenges. WikiDBGraph thus serves as a crucial benchmark to measure this gap and guide future research into more effective and scalable collaborative learning paradigms.

%% file: tech-report.tex
\begin{center}
{\Large\bfseries Technical Report}\\[0.6em]
\end{center}

\begin{revblock}

\section{Extension to GitTables}\label{sec:gittables_ext}

To evaluate the generalization of our contrastive framework beyond WikiDBGraph, we extend our analysis to GitTables~\cite{hulsebos2023gittables}, a corpus of heterogeneous web tables lacking explicit schema mappings. To adapt to this unsupervised setting, we introduce a self-supervised \textit{table partition matching} task.

\paragraph{Task Formulation}
Unlike WikiDBGraph, which utilizes shared Topic IDs for supervision, we generate ground truth via synthetic partitioning. We construct training triplets $(x_a, x_p, x_n)$ where the anchor $x_a$ and positive $x_p$ are disjoint partitions of the same table $T$, while $x_n$ is a partition from a randomly sampled table $T_j (j \neq i)$. We employ two partitioning strategies to simulate collaborative learning scenarios:
(1) \textit{Vertical Partitioning}, which splits columns to mimic feature-partitioned silos; and 
(2) \textit{Horizontal Partitioning}, which splits rows to mimic sample-partitioned silos.

\paragraph{Experimental Setup}
We use the partition \texttt{whole\_tables\_licensed} of GitTables~\cite{hulsebos2023gittables}, applying strict filtering criteria (minimum 50 rows, minimum 6 columns) to ensure sufficient content for partitioning. This results in 46,325 valid tables, from which we generate 72,685 triplets. The data is partitioned into an 80/10/10 split.

Similar to Section~\ref{subsec:approach}, input tables are serialized using both schema and sampled data: \texttt{[col\_name]: val1, val2, ...}, with a sample size of 3 values per column and a maximum context of 50 columns. We fine-tune the \texttt{BGE-M3} model using InfoNCE loss objective with cosine similarity according to Eq.~\ref{eq:infonce_loss}. Training is conducted for 10 epochs using the AdamW optimizer with a learning rate of $1\times10^{-5}$ and a batch size of 32. During training, we sample 2 negatives per anchor.

\begin{table}[ht]
    \centering
    \caption{GitTables relation prediction performance: Original (BGE-M3) vs contrastive fine-tuned like WikiDBGraph. Best in \textbf{bold}.}
    \label{tab:gittables-results}
    \begin{tabular}{lcc}
        \toprule
        \textbf{Metric} & \textbf{Original} & \textbf{Contrastive Learning} \\
        \midrule
        Accuracy & 0.7454±0.0009 & \textbf{0.9712±0.0010} \\
        Precision & 0.7666±0.0115 & \textbf{0.9533±0.0014} \\
        Recall & 0.7067±0.0195 & \textbf{0.9909±0.0006} \\
        F1 & 0.7351±0.0056 & \textbf{0.9718±0.0009} \\
        AUC-ROC & 0.8297±0.0008 & \textbf{0.9888±0.0006} \\
        \bottomrule
    \end{tabular}
\end{table}

\paragraph{Results}
Evaluation is performed on the held-out test set (7,269 triplets) using a 1:1 balanced protocol across 5 random seeds. The results are detailed in Table~\ref{tab:gittables-results}. Contrastive fine-tuning yields substantial improvements across all metrics compared to the pretrained baseline. Specifically, Accuracy increases from 74.54\% to 97.12\%, and the F1-score jumps from 0.7351 to 0.9718. The near-perfect AUC-ROC (0.9888) and Recall (0.9909) confirm that our objective effectively captures table provenance and semantic coherence even without explicit supervision. The result concludes that our approach can also generalize to other table corpora. 

These results confirm that our contrastive objective generalizes beyond WikiDBGraph and captures table- or database-level provenance across diverse corpora. Crucially, the framework remains effective even when explicit supervision signals such as TIDs are unavailable, supporting its applicability to broader collections of tables and databases.

\section{Performance Relation to Database Size}\label{sec:perf_vs_size}

To investigate the robustness of collaborative learning across varying data scales, we analyze the performance improvement ($F1_{\text{FedAvg}} - F1_{\text{Solo}}$) relative to the local dataset size in Fig.~\ref{fig:size_vs_gain_fedavg}. This analysis reveals two critical findings regarding the relationship between data volume and graph utility.

First, when employing naive string-based alignment (blue trend), we observe a distinct transition in graph utility relative to data scale: while small datasets ($<10^2$ rows) benefit from external signals due to the ``data starvation'' effect, larger datasets ($>10^3$ rows) suffer diminishing returns and potential negative transfer, as misaligned external features become noise that destabilizes already robust local models.

Second, employing semantic-aware alignment (DeepJoin~\cite{dong2023deepjoin}, orange trend) effectively resolves this regression risk. The DeepJoin trend line remains consistently positive even for the top quartile of dataset sizes, with 77.4\% of tasks exhibiting neutral or positive transfer. This confirms that the graph structure remains valuable for larger databases provided that preprocessing effectively resolves semantic correspondence, thereby attributing the regression risk to alignment quality rather than the intrinsic quality of the database connections.

\begin{figure}[ht]
    \centering
    \includegraphics[width=0.95\linewidth]{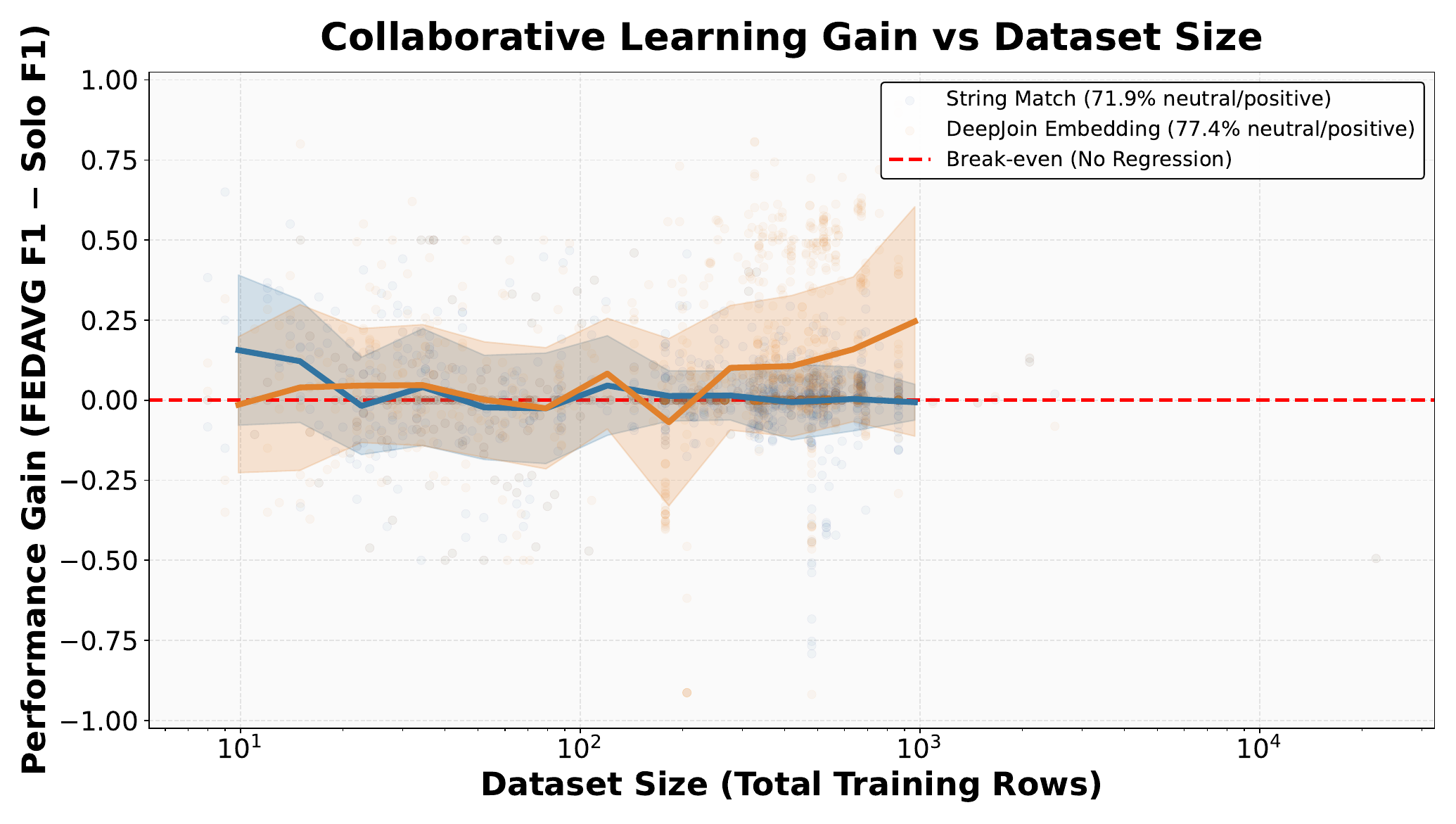}
    \caption{Impact of dataset size on collaborative learning gain. The trend line represents the mean performance with variance indicated by the shaded region. Individual dots correspond to the performance of distinct database pairs.}
    \label{fig:size_vs_gain_fedavg}
\end{figure}

\end{revblock}

\section{Discussion}
\label{sec:discussion}

\input{sections/discussion}

\section{Future Directions}
\label{sec:future}
\input{sections/future}

%% file: sections/discussion.tex
\begin{table*}[tb]
    \centering
    \caption{Summary of key findings and implications for practical Collaborative Learning (CL)}
    \label{tab:key_findings}
    \begin{tabular}{p{6.5cm} p{7.5cm} p{2.5cm}}
    \toprule
    \textbf{Finding / Implication for Collaborative Learning} & \textbf{Supporting Evidence (This Work)} & \textbf{Reference(s)} \\
    \midrule
    CL overall outperforms \textit{Solo} training on real-world databases. & CL algorithms such as \textit{FedAvg}, \textit{FedProx}, and \textit{FedOV} show improvements over \textit{Solo} in both the overall analysis (Fig.~\ref{fig:fl_performance_scores}) and the case studies (Tables~\ref{tab:horizontal_fl_comparison}, \ref{tab:vertical_fl_comparison}). & Aligns with \cite{li2022federated,mcmahan2017communication} \\
    \midrule
    Advanced CL algorithms do not consistently outperform \textit{FedAvg} on non-IID data. & Relational schemas typically contain non-IID data, producing performance discrepancies across clients, as observed in the overall analysis (Figs.~\ref{fig:fl_performance_scores}) and the case study (Table~\ref{tab:horizontal_fl_comparison}). & Aligns with \cite{li2022federated} \\
    \midrule
    Databases in CL form highly interconnected networks with a long-tailed degree distribution; links should neither be ignored nor treated uniformly. & WikiDBGraph exhibits a long-tailed node-degree distribution (Fig.~\ref{fig:degree-distribution}), with low median but high maximum degrees (Table~\ref{tab:graph-stats}). & Similar to social/web networks \cite{enders2008long,anderson2007long}; newly reported for CL \\
    \midrule
    New CL paradigms are needed for relational databases to avoid computationally infeasible full-table joins. & The ``AllJoinSize'' is infeasible on average ($9.5\times10^{16}$ rows; Table~\ref{tab:node-props}), making single-table approaches impractical. & Novel finding \\
    \midrule
    CL algorithms should handle \textbf{hybrid} cases with partial instance and feature overlap, beyond ideal alignment assumptions. & Partial data and schema overlap are evidenced by the low \texttt{OverlapRatio} and \texttt{JaccardColumn} (Table~\ref{tab:edge-props}) and highlighted in the hybrid case study (Section~\ref{subsec:case-study-hybrid}). & Novel finding \\
    \midrule
    Schema matching and entity resolution are critical, unresolved bottlenecks for real-world CL. & Manual schema/feature alignment in case studies (Section~\ref{sec:case_study}) yields strong \textit{Combined} performance, whereas in automated evaluation only a subset of CL tasks benefit and average improvements are small for both \textit{Combined} and CL algorithms. & Novel finding \\
    \bottomrule
    \end{tabular}
\end{table*}

This section summarizes key findings and implications for practical CL tasks derived from WikiDBGraph; a concise overview appears in Table~\ref{tab:key_findings}. Some findings are consistent with established studies~\cite{li2022federated,mcmahan2017communication}, others parallel results reported in adjacent research areas~\cite{enders2008long,anderson2007long} but not previously for CL, and many are novel to our analysis. Collectively, these results highlight the unique complexities of real-world data collaboration and motivate further investigation.

\paragraph{Effectiveness of Collaborative Learning}
Both the overall analysis (Fig.~\ref{fig:fl_performance_scores}) and the case studies (Tables~\ref{tab:horizontal_fl_comparison}, \ref{tab:vertical_fl_comparison}) show that CL algorithms generally outperform \textit{Solo} training on real-world databases. However, advanced CL methods such as \textit{FedProx}~\cite{li2020federated} do not consistently improve over simpler methods like \textit{FedAvg}~\cite{mcmahan2017communication} on non-IID data. Both observations align with prior CL benchmarks~\cite{li2022federated}, supporting WikiDBGraph as a valid benchmark for evaluating CL effectiveness.

\paragraph{Interconnection of Collaborative Learning}
Databases in WikiDBGraph form highly interconnected networks with a long-tailed degree distribution. As evidenced by Fig.~\ref{fig:degree-distribution} and Table~\ref{tab:graph-stats}, some nodes have high degree while many have low degree. Thus, most existing CL approaches \cite{li2023fedtree,diao2023towards} that ignore links or treat all nodes uniformly are suboptimal. This property mirrors patterns in other real-world networks (e.g., social and web graphs)~\cite{enders2008long,anderson2007long} and underscores the need for graph-aware CL frameworks.

\paragraph{Infeasibility of Full Table Joins.}
The average \texttt{AllJoinSize}---the estimated row count from joining all tables within a database---is $9\times10^{16}$ (Table~\ref{tab:node-props}), rendering single-table approaches computationally infeasible. Consequently, CL methods that require full table joins are impractical for large real-world databases. Although such databases may be less common, they are often highly valuable due to their scale, calling for CL frameworks that operate directly on relational structures rather than a single prejoined table.

\paragraph{Hybrid Alignment in Collaborative Learning}
Databases in WikiDBGraph are neither purely horizontally nor purely vertically aligned; instead, they exhibit \textit{hybrid} alignment. This is supported by low \texttt{OverlapRatio} and \texttt{JaccardColumn} values in Table~\ref{tab:edge-props} and is further elaborated in the hybrid case study (Section~\ref{subsec:case-study-hybrid}). As a result, CL approaches that assume ideal horizontal or vertical alignment cannot handle many real-world cases, motivating methods that explicitly accommodate hybrid alignment.

%% file: sections/future.tex
This section presents three main opportunities for data management research on WikiDBGraph.

\noindent\textbf{Schema Matching:} Analyses on WikiDBGraph indicate strong potential for schema matching~\cite{bernstein2011generic}, as identifying correct feature correspondences is crucial for learning accuracy. The dataset also provides ground truth for schema matching not only between two databases but also across multiple databases with complex relationships. WikiDBGraph surpasses prior datasets such as GitSchema~\cite{dohmen2024gitschemas} in scale and offers rich data beyond schemas alone. Its data distributions can further aid column alignment, enabling data-aware schema matching approaches. 

\noindent\textbf{Record Linkage and Data Mining:} The prevalence of unaligned data suggests that record linkage~\cite{vatsalan2017privacy} (a.k.a.\ entity alignment~\cite{zhao2020experimental}) is inherently noisy~\cite{nock2021impact}. Improving alignment in isolation, without feedback from downstream data mining, hinders task-aware alignment. A critical direction is to mine data under inaccurate or incomplete row alignment, which typically calls for a coupled data-mining framework~\cite{wu2022coupled}. 

\noindent\textbf{Graph Mining:} To incorporate databases in CL that cannot be directly joined, graph mining~\cite{rehman2012graph} techniques must be extended to support CL over database graphs---tables connected by foreign keys. This join challenge calls for new graph-based CL methods beyond current table-centric approaches.